\documentclass[a4,10pt]{article}
\usepackage{graphicx}
\usepackage{natbib}
\usepackage{psfrag}
\usepackage{bm}
\usepackage{color}
\usepackage{amsmath}
\usepackage{amssymb}
\usepackage{amsfonts}
\usepackage{fancyhdr}
\usepackage{float}

\usepackage{wasysym}
\usepackage{multirow}
\usepackage{array}
\usepackage{mathrsfs}
\newcommand\etal{\mbox{\textit{et al.}}}

\newsavebox{\astrutbox}
\sbox{\astrutbox}{\rule[-5pt]{0pt}{20pt}}

\newcommand{\BEQ}{\begin{equation}}
\newcommand{\EEQ}{\end{equation}}
\newcommand{\BER}{\begin{eqnarray}}
\newcommand{\EER}{\end{eqnarray}}
\newcommand{\BEQN}{\begin{equation*}}
\newcommand{\EEQN}{\end{equation*}}
\newcommand{\BERN}{\begin{eqnarray*}}
\newcommand{\EERN}{\end{eqnarray*}}

\makeatletter
\newcommand*{\rom}[1]{\expandafter\@slowromancap\romannumeral #1@}
\makeatother

\begin{document}

\fancyhf{}
\pagestyle{fancy}
\rhead{\textit{DNS of MATUR}}
\lhead{\textit{L.\ Chen \etal}}

\title{Direct numerical simulation of quasi-two-\\dimensional MHD turbulent shear flows}

\author{
Long Chen$^{1}$, 
Alban Poth\'{e}rat$^{2}$, 
Ming-Jiu Ni$^{1}$
and Ren\'{e} Moreau$^{3}$\\

$^1$School of Engineering Science, University of Chinese Academy of Sciences,\\ Beijing 101408, China, mjni@ucas.ac.cn\\
$^2$Fluid and Complex Systems Research Centre,\\ Coventry University, Coventry CV15FB, United Kingdom\\
$^{3}$Universit\'{e} de Grenoble, Laboratoire SIMAP, Groupe \\
EPM, BP 75, 38402 Saint Martin d'H\`{e}res, France
}

%
\maketitle
%
\begin{abstract}
High-resolution direct numerical simulations (DNS) have been performed to study the turbulent shear flow of an electrically conducting fluid in a cylindrical container. The flow is driven by an annular azimuthal Lorentz force induced by the interaction between the radial electric currents ($I$) injected through a large number of small electrodes placed in the bottom wall and the magnetic field ($B$) imposed in the axial direction. All the numerical parameters, including the geometry of the container, the value of the external electric currents and the strength of the magnetic fields, are set to be in line with the experiment performed by \citet{Messadek2002} (J. Fluid Mech. \textbf{456}, 137-159).
Firstly, maintaining the Hartmann layers in the laminar regime, three dimensional simulations are carried out to reproduce some of the experimental observations, such as the global angular momentum and the velocity profiles. In this regime, the variation laws of the wall shear stresses, the energy spectra and the visualizations of the flow structures near the side wall indicate the presence of separation or turbulence within the side wall layers, even though the current injection electrodes are far from the side wall. Furthermore, a parametric analysis of the flow also reveals that the Ekman recirculations have significant influence on the vortex size, the free shear layer, and the global dissipation. Second, we recover the scaling laws of the cutoff scale that separate the large quasi-two-dimensional scales from the small three-dimensional ones (\citet{Sommeria1982} J. Fluid Mech. \textbf{118}, 507-518), and thus establish their validity in sheared MHD turbulence. Furthermore, we find that three-componentality are and the three-dimensionality appear concurrently and that both the two-dimensional cutoff frequency and the mean energy associated to the axial component of velocity scale with $N_t$, respectively as $0.063N_t^{0.37}$ and $0.126 N_t^{-0.92}$.
\end{abstract}

\section{\label{first:level1}Introduction}

In this paper, we apply the three-dimensional (3D) direct numerical simulations (DNS) to study the flow of an electrically conducting incompressible fluid in a cylindrical container, popularly known as MATUR (MAgnetohydrodynamics TURbulence), designed to  investigate the quasi-two-dimensional (Q2D) turbulence \citep{Alboussiere1999} in the presence of external magnetic fields (MHD). MATUR is shown in Fig. \ref{model}, where the  magnetic field is applied along the axial direction. For such a laboratory scale configuration, the magnetic Reynolds number is much smaller than unity (here, $R_{m}= \mu_{m}\sigma U_{0} L\approx 0.007$, where $\mu_{m}$ denotes the magnetic permeability of vacuum, $\sigma$ is the electrical conductivity, $U_{0}=0.1$ m/s and $L=0.01$ m are the typical characteristic velocity and the characteristic length scale, respectively), then the induced magnetic field $\textit{\textbf{b}}$ is much smaller than the imposed one $\textit{\textbf{B}}$ ($\textit{\textbf{b}}\sim R_{m}\textit{\textbf{B}} \ll \textit{\textbf{B}}$), and the Lorentz force acting on the flow is obtained as $\textit{\textbf{F}}_{Lorentz}=\textit{\textbf{j}}\times\textit{\textbf{B}}$, where $\textit{\textbf{j}}$ denotes the electric current density \citep{Roberts1967}. If the magnetic field is strong enough,
velocity variations along the field lines are damped by the Lorentz force and the flow tends to be Q2D.

This tendency was observed in several experiments. \citet{Kolesnikov1974} and \citet{Davidson1997} stated that this evolution towards a quasi-two-dimensional regime was a consequence of the invariance of the angular momentum component parallel
to the magnetic field, when its perpendicular components decay exponentially
($ \sim e^{\sigma B^{2}t/ \rho} $). \citet{Eckert2001} conducted an experiment on MHD turbulence in a sodium channel flow exposed to a transverse magnetic field, and the measured turbulence intensity and energy spectra were found to exhibit a spectral slope varying with the magnetic interaction parameter ($N=\sigma B^{2}L/\rho U_{0}$) from a $k^{-5/3}$ law for $N \leq 1$ and to a the exponent a minimum of  $-4$ for $N\simeq 120$. For the MHD turbulent shear flows, \citet{Kljukin1989} performed the very first experiments, in which the mean velocity distribution and correlations of velocity fluctuations were provided, but data concerning the energy spectra and the development of coherent structures fed by the energy transfer towards the large scales was not available.

In order to better understand the elementary properties of the Q2D turbulent shear flows and check the validation of theoretical work (\citet{Sommeria1982}), \citet{Alboussiere1999} and  \citet{Potherat2000} carried out experimental and theoretical studies on the Q2D turbulent shear flows, where the transport of a scalar quantity and the free surface effect were considered, using the MATUR equipment. \citet{Messadek2002} further provided experiment data on the MHD turbulent shear flows in a wide range of $Ha$ and $Re$,
and highlighted the important role of the Hartmann layers where the Joule effect and viscosity dissipate most of the kinetic energy. Recently, \citet{Stelzer2015b} built a new experimental  device called ZUCCHINI (ZUrich Cylindrical CHannel INstability Investigation), which, as MATUR, featured a free shear layer at the edge of inner disk electrode. Combining it with finite element simulations (based on a 2D axisymmetric model), they studied the instabilities of the free shear layer and identified several flow regimes characterized by the nature of the instabilities of the Kelvin-Helmholtz type \citep{Stelzer2015a}. Based on the FLOWCUBE platform, a more homogeneous type of turbulence between Hartmann walls was produced from the destabilisation of vortex arrays (\citet{KLEIN2010}; \citet{Potherat2014}; \citet{Baker2018}). These authors focused on the transition between 3D and Q2D turbulence. In particular, the cut-off length scale $\widehat{l}^{c}_{\perp}$ ($\sim N^{1/3}_{t}$, where 
\begin{equation}
N_{t}=\frac{\sigma B^2a}{\rho U_0}\left(\frac{a}{L}\right)^2
\label{eq:nt}
\end{equation}
%
 is the true interaction parameter) first theorised by \cite{Sommeria1982}, that separate 3D from Q2D fluctuations were obtained experimentally, as well as the evidences of inverse and direct energy cascades in 3D magnetohydrodynamic turbulence.

However, a major disadvantage of experimental approaches is that the liquid metal used for their high electrical conductivity is non-transparent. Although the velocity fields can be measured  by ultrasonic Doppler velocimetry  or
potential probe techniques, more complete information, e.g  the distribution of the flow fields and the electromagnetic quantities, are rather difficult to obtain. Therefore,  numerical simulations, which can complement  the experimental measurements, have been developed recently to study  MHD turbulence. Taking advantage of the Q2D property of the MHD flows in case of high $N$ and $Ha$, several simplified effective 2D models have been developed by averaging the full Navier-Stokes
equations along the direction of the magnetic fields. The advantages of using these 2D models are evident, not only to save the costs compared to a full 3D numerical approach, but also to provide accurate results where 3D numerical solutions cannot fully resolve the boundary layer in case of high $Ha$. \citet{Sommeria1982} derived a two-dimensional model (denoted as SM82 hereafter) based on the simple exponential profile of Hartmann layers. It gave good
results in the flow regime where inertia is small but failed to describe flows where strong rotation induces
secondary flows, such as Ekman pumping.  The 2D model developed by \citet{Potherat2000} (denoted as PSM hereafter), accounting for some 3D effects, gave more accurate prediction in the Q2D flows. With PSM, both of the velocity profiles and the global angular momentum measurements from MATUR  \citep{Potherat2005} were reproduced, and it was proved that the local and global Ekman recirculations altered the shape of the flow significantly as well as the global dissipation. However, both the SM82 model and the PSM model break down if the Hartmann layer becomes turbulent, where the flow may still remain quasi-two-dimensional but the boundary layer friction is altered.  \citet{Potherat2011} established an alternative shallow water model specifically for this case, and recovered experimentally measured velocity profiles and global momentum in this regime.

In mainly azimuthal flows such as in a toroidal containers, the dynamics of the side wall layer and the free shear layer near the injected electrodes on the flow is complex because of rotation effects. Even when the Hartmann layers are stable, significant flow alterations may occur, including non-trivial 3D effects \citep{Tabeling1981}, which could not be observed easily in experiments or with any Q2D model.
It has been proven that
turbulence may remain localized in a layer near the outer cylinder wall prior to transition happening in Hartmann layers as indicated by \citet{zhao2012},  who conducted a series of 3D DNS of MHD turbulence flows in a toroidal duct. In addition, for cases with lower values of Hartmann number and higher values of Reynolds number,  three-dimensionality would be more pronounced, even within the $Hartmann-B\ddot{o}dewadt$ layers, which have been studied theoretically by \citet{Davidson2002} and \citet{Moresco2003}. All of these discoveries encourage us to perform 3D DNS on the flows in MATUR configuration (corresponding to the realistic experiment of \citet{Messadek2002}). Besides reproducing the results obtained in the experiments, theories and Q2D simulations, we focus on answering the following questions in the present work.
\begin{enumerate}
\item Does the separation or turbulence emerge within the side wall layer  when the electrodes are far away from the side wall and while the Hartmann layer remains laminar?
\item What causes the angular momentum dissipation in regimes where the Hartmann layer is laminar?
\item How much and what type of three-dimensionality subsists in sheared MHD turbulence at high $Ha$? In particular,
\begin{enumerate}
\item How much energy subsists in the secondary flow?
\item Is there a cutoff lengthscale between quasi-2D and 3D lengthscales in sheared turbulence too?
\end{enumerate}
\end{enumerate}
However, two factors restrict the investigated range of Reynolds number and Hartmann number in the present DNS
studies. One is the computing resource, because very fine grids are required to capture the small-scale turbulent structure and to resolve the thin Hartmann
boundary layers. Another is the lack of robust computational schemes capable
of dealing with nonlinear unsteady high-$Ha$ flows. In particular, when non-orthogonal grids are used, extra non-orthogonal correction schemes are required. By applying Large eddy simulations \citep{Kobayashia2006,Kobayashia2008}, $Re$ could be
somewhat increased, but the resolution requirements for the Hartmann layers remained
essentially the same as those in DNS, since no reliably accurate wall function models were
known for the case of turbulent flows. Here, the problem of inadequate computational resources
is overcome by employing massively parallel computing. As for the numerical method, we apply the
finite volume method based on the consistent and conservative scheme developed by \citet{Ni2007}, which can be used to accurately simulate MHD flows at a high Hartmann number. Therefore, the Hartmann number is allowed to vary from $55$ to $792$ (magnetic fields change from $\rm 0.2083$ $\textrm{T}$ to $\rm 3$ $\textrm{T}$) while the Reynolds number also varies from $4792$ to $31944$ (total current density change from $\rm 3$ $\textrm{A}$ to $\rm 20$ $\textrm{A}$). In such parameter spaces, turbulence is well established  while the Hartmann layers remain laminar. Simulations are performed on the full three-dimensional domain and, for comparison with previous work, with  the PSM model in the 2D-average plane.

The layout of the paper is as follows: In $\S$ \ref{sec:2}, a short description of the physical model underlying this
work  and the flow conditions in the MATUR cell are given. Particular attention is given to the modifications dealing with the electrical conductive wall and the injected current density. The numerical algorithm  and the detailed computational grid study are also presented in this part.  The main properties of the MHD turbulence are described and discussed in $\S$ \ref{sec:3}, including the general aspect of the flow, the secondary flows, the properties of the free shear layer and side wall layer, the global angular momentum as well as three dimensionality. Finally, we offer concluding remarks in $\S$ \ref{sec:4}.
\section{Problem statement and formulation}\label{sec:2}
\subsection{Flow configuration and mathematical formulation}
\begin{figure}
  \centering
  \includegraphics[width=0.7\textwidth]{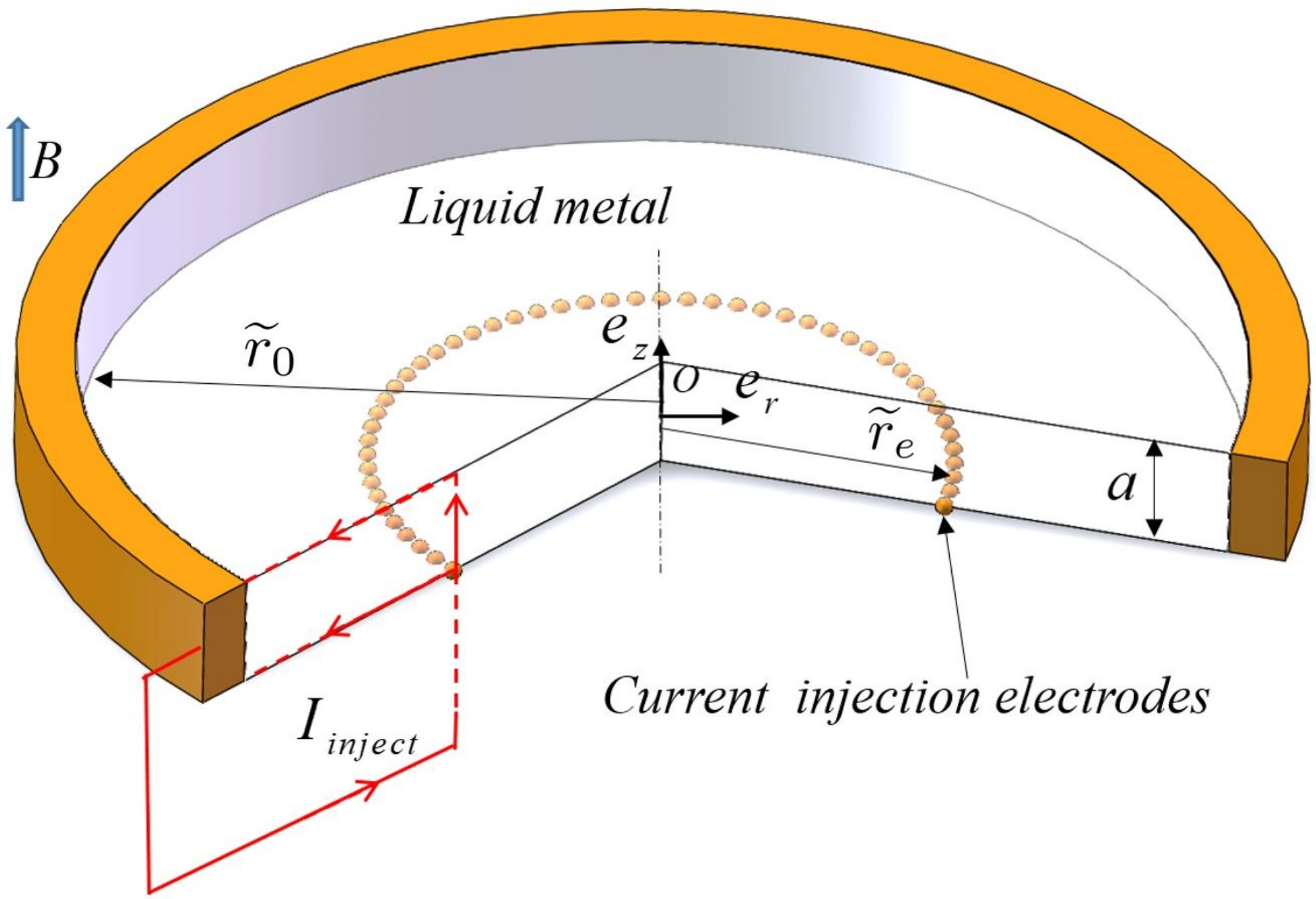}
  \caption{\label{model}Sketch of the experimental set up. A typical electric
circuit including one of the point-electrodes mounted flush at the bottom
Hartmann layer is represented. }
\end{figure}

The physical model of MATUR is shown in Fig.\ref{model}.
It is a cylindrical container (radius $\widetilde{r}_{0}=\rm 0.11$ m, depth $a= \rm 0.01$ m, with $\widetilde{}$ distinguishing the dimensional quantity from their dimensionless counterpart), in which the bottom and the upper walls are electrically insulating while the vertical walls are conducting.
Electric currents are injected at the bottom of the container through a large number of point-electrodes spread along a
 concentric ring parallel with the vertical wall. As indicated by \citet{Messadek2002}, a continuous electrode ring would induce a strong local
damping in the flows, and thus, a series of discrete point-electrodes are positioned to reduce this unwanted effect.  The concentric circles are located at $\widetilde{r}_{e}=\rm0.054$ m
 or $\widetilde{r}_{e}=\rm0.093$ m, respectively. In the present study, only $\widetilde{r}_{e}=\rm0.054$ m is considered. The container is filled with mercury, and exposed to a constant homogeneous vertical magnetic field parallel to the axis of MATUR. The injected currents leave the fluid through the vertical wall,
 to induce radial electric current that give rise to an azimuthal force on the fluid in the annulus
 between the electrode circle and the outer wall.
The material properties of the fluid at room temperature, such as the mass density $\rho$, the kinematic viscosity $\nu$ and
the electrical conductivity $\sigma$, are assumed constant ($\rho=1.3529\times10^{4}$ kg/m$^{3}$, $\nu=1.1257\times10^{-7}$ m$^2$/s and $\sigma=1.055\times10^{6}$ S/m).  An external homogeneous magnetic field of amplitude $B$ is applied along the axial direction. At a low magnetic Reynolds number, the full system of the induction equation and the Navier-Stokes equations for an incompressible fluid can be approximated to the first order $\mathcal O(Rm)$. Thus, the  non-dimensional magnetohydrodynamic equations governing the flow can be written as \citep{Roberts1967}
\begin{equation}
\nabla \cdot \textbf{\textit{v}} =0,
\end{equation}
\begin{equation}
  \frac{\partial\textbf{\textit{v}}}{\partial t}+(\textbf{\textit{v}}\cdot\nabla)\textbf{\textit{v}}=-\nabla p+\frac{1}{Re}\Delta\textbf{\textit{v}}+N(\textbf{\textit{j}}\times\textbf{\textit{e}}_{z}),
   \label{Eq:momentumEq}
\end{equation}
\begin{equation}
  \textbf{\textit{j}}=-\nabla\varphi+\textbf{\textit{v}}\times\textbf{\textit{e}}_{z},
\end{equation}
\begin{equation}
  \nabla \cdot \textbf{\textit{j}} =0.
\end{equation}
where the variables $\textit{\textbf{j}}$, $\varphi$, $\textbf{\textit{v}}$, $p$ denote the current density, the electric potential, the velocity and the pressure, respectively. Here, lengths are scaled by $a$, the  velocity by a scale $U_{0}$ to be specified shortly, and the  current by $ \sigma B U_{0}$. The typical scales for the other variables are as follows: $\rho U_{0}^{2}$ for the pressure, $U_{0} B a$ for the electrical potential, $a/U_{0}$ for time. 
The Hartmann number $Ha$ and the interaction parameter $N$ are defined as
\begin{equation}
 Ha=B a\sqrt{\frac{\sigma}{\rho\nu}}, \quad N= \frac{\sigma B^{2} a}{\rho U_{0}},
\end{equation}
and the Reynolds number is given as $Re=Ha^{2}/N$. In the present work, all these non-dimensional numbers are based on the thickness of the container $a$. For this experiment, an approximate azimuthal velocity can be derived from the theory in \citet{Sommeria1982}. Indeed, \citet{Potherat2000} derived an approximate expression for the $z$-averaged azimuthal velocity in the inviscid laminar, axisymmetric Q2D regime, using a Dirac delta
function centred at the electrodes $r=r_{e}$ to describe the injected current $j_{W}$, where the integral is equal to the total injected current $I$: $j_{W}=I/2\pi r_{e}\delta(r-r_{e})$.
\begin{equation}
\left\{
\begin{array}{lr}
    U_\theta^{\rm SM82}(r)=\frac{I}{2\pi  r \sqrt{\sigma\rho\nu}}, \quad r_{e}<r<r_{0}, & \\
    U_\theta^{\rm SM82}(r)= 0, \quad r\leq r_{e}. &
\end{array}
\right.
\label{Eq:Uteta1}
\end{equation}
and $U_r^{\rm SM82}=0$. 
Based on this, we choose the velocity scale as $U_{0}=\frac{I}{2\pi  r_{0}\sqrt{\sigma\rho\nu}}$.
Finally, velocity fluctuations are defined as  $\textbf{\textit{v}}'=\textbf{\textit{v}}-\langle \textbf{\textit{v}}\rangle$.

The boundary conditions for $\textbf{\textit{v}}$ and $\varphi$ are as follows. For the velocity,
we apply standard no-slip conditions at all walls. As for the electric potential, we impose perfectly
conducting side walls and perfectly insulating Hartmann walls except for the locations where the electrodes are located, \emph{i.e.}: at the top wall,
\begin{equation}
\textbf{\textit{v}} =0, \quad \partial_{z}\varphi=0 \quad {\rm at} \quad z=1 \quad {\rm (top \quad wall)},
\end{equation}
at the surface of point electrodes located at the bottom wall:
\begin{equation}
  \textbf{\textit{v}} =0, \quad \partial_{z}\varphi=\frac{I}{A_{re}\sigma^{2} BU_{0}} \quad {\rm at} \quad z=0
   \label{Eq:currentinject}
\end{equation}
at bottom wall, outside point electrode surfaces:
\begin{equation}
 \textbf{\textit{v}} =0, \quad \partial_{z}\varphi=0 \quad {\rm at} \quad z=0 
\end{equation}
at the lateral wall (BC1),
\begin{equation}
  \textbf{\textit{v}} =0, \quad \varphi=0 \quad {\rm at} \quad r=r_0 
  \label{Eq:currentout}
\end{equation}
%
Here, $128$  points-electrodes ($0.001$ m diameter and $0.00165$ m apart) are uniformly distributed along a circle located at $r_e$  at the bottom wall. The surface of each point-electrode contains 102 cells with those on the periphery cut by the electrode edge. The area $A_{re}$, which appears in the boundary condition of (\ref{Eq:currentinject}), is the total area of the small electrodes.  
Note that we do not need an extra forcing term in (\ref{Eq:momentumEq}) to drive the flow because the injected current from the bottom electrodes  interacts with the magnetic field, and generates Lorentz force that drives the flow. The current circulate inside the mercury, as shown in figure \ref{model}, and leaves the domain at the vertical wall. Moreover, because of the perfectly conducting properties of the wall, the electrical potential across the wall can be regarded as a constant, and set to zero in (\ref{Eq:currentout}).

\subsection{Numerical algorithm and validation}
The direct numerical simulations of the governing equations are performed based on the finite volume approach.
For the pressure$-$velocity coupling, a second-order temporal accurate pressure-correction algorithm has been used. Based on a consistent and conservative scheme \citep{Ni2007}, the electrical potential Poisson  equation is then solved to obtain $\varphi$. The detailed process within each time step could be split into: a) obtain a predicted velocity by solving the
momentum equation with pressure  from the previous iteration;
b) calculate the predicted velocity fluxes which are used as the source term
of the pressure difference Poisson equation, which will be solved to obtain the pressure difference; then apply the updated pressure difference to update the velocity and pressure;  c) solve the Poisson equation for the electric potential to get the electrical potential, which is used to calculate the current density fluxes on cell faces with the consistent and conservative scheme. Then the current density at each cell centre is reconstructed through a conservative interpolation $\textbf{\textit{j}} = \nabla\cdot(\textbf{\textit{jr}})$ with $\textbf{\textit{r}}$ the position vector; d) calculate the Lorentz force $\textit{\textbf{F}}_{Lorentz} = \textit{\textbf{j}}\times \textit{\textbf{e}}_{z}$ at the cell centre based on the reconstructed current density, which is used as the source term of momentum equation of next time step. Step (b) is  iterated 3 times before solving the electrical potential Poisson  equation.

The central scheme  is applied for all convective-term approximations. All inviscid terms and
the pressure gradient are approximated with a second-order accuracy. A second order implicit Euler method is used for time integration.
In order to guarantee a robust solution for unsteady flows and  make the temporal cutoff frequency match the spatial cutoff frequency, the present simulations are run with a constant time step
which satisfies the Courant-Friedrichs-Lewy condition.

 A Preconditioned Bi-Conjugate Gradient solver
(PBiCG) applicable to asymmetric matrices has been used for the
solution of the velocity-pressure coupling equation, together with a Diagonal-Incomplete LU
(DILU) decomposition for preconditioning. The pre-conditioned conjugate gradient (PCG) iterative solver with a diagonal incomplete Cholesky (DIC) pre-conditioner, which deals with symmetric matrices, has been applied
for the solution of pressure and electric potential equations. Note that for all the iterative solutions,  velocity, pressure and electric potential, a constant convergence criterion of $10^{-6}$ is used.

  \begin{figure}
  \centering
    \includegraphics[width=0.45\textwidth]{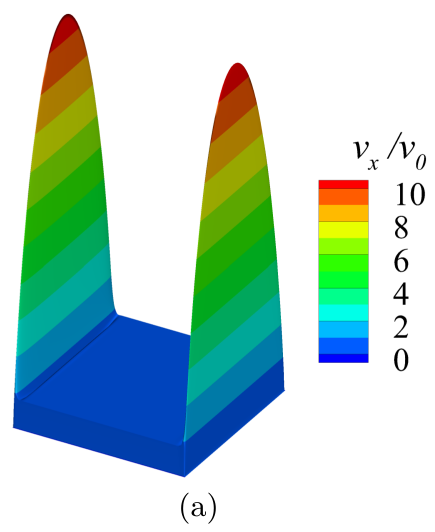}
    \includegraphics[width=0.45\textwidth]{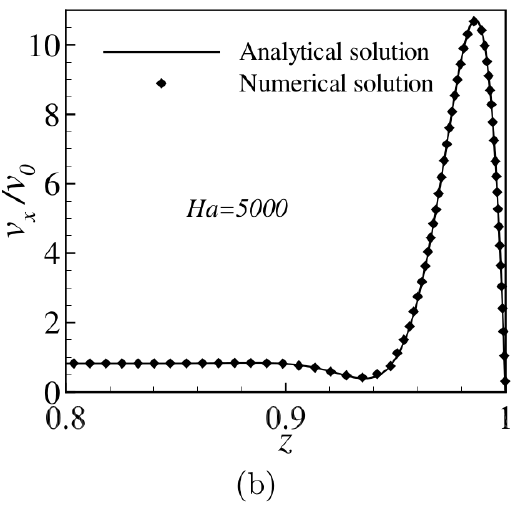}
  \caption{\label{hunt-compare} The typical "M-shape" velocity profile of Hunt's case (a) and the comparison of the numerical result with Hunt’s analytical solution (b) for $Ha = 5000$. 
}
\end{figure}

  \begin{figure}
  \centering
  
  \includegraphics[width=0.45\textwidth]{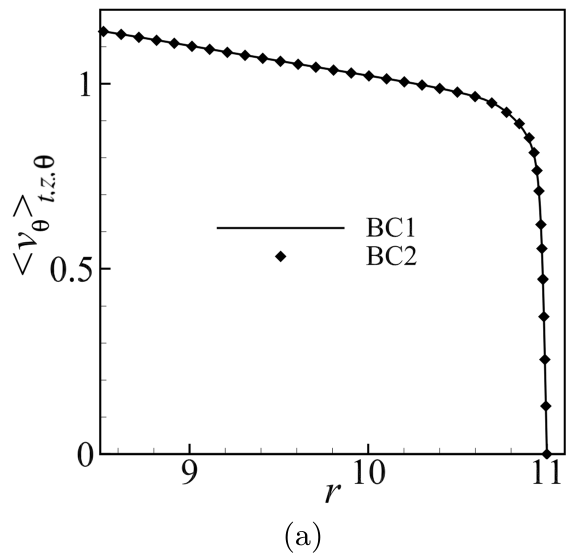}
    \includegraphics[width=0.45\textwidth]{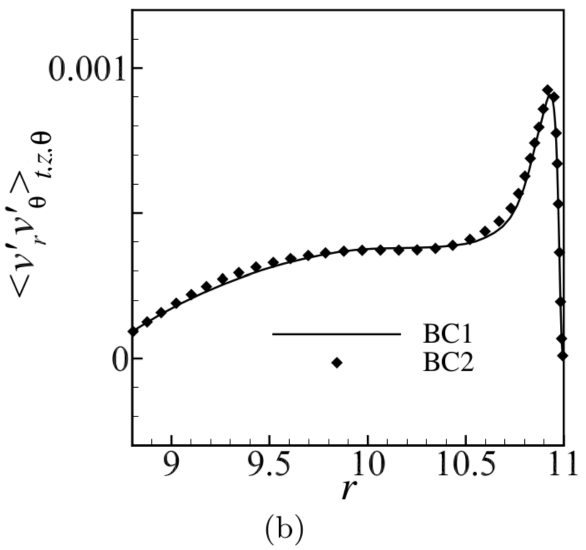}
  \caption{\label{bc-compare} The radial distribution of the mean azimuthal velocity (a) and the rms of velocity fluctuations (b) for BC1 and BC2 at $Ha = 66$ and $Re= 15972$.} 
\end{figure}

In order to verify the accuracy of our numerical code,  the classic Hunt's flow has been simulated for comparison with the analytical solution (Hunt 1965). Herein, the parameters are set to $Re=100$, $Ha=5000$, and the  conductance ratio of $C_{w}=\sigma_{w}t_{w}/\sigma_{f}L_{f}$ is set to 0.01.
As illustrated in figure \ref{hunt-compare}, the velocity matches well with Hunt's analytical result, especially within the thin boundary layer
from $z = 0.8$ to $z = 1$. Moreover, the calculated pressure gradient matches
well with the analytical pressure gradient,  with a relative discrepancy lower than $0.08\%$ based on an error estimation of $|\frac{(\nabla p)_{Anal}-(\nabla p)_{Numr}}{(\nabla p)_{Anal}}|$. 

In addition, it should be noted that the boundary condition of a perfectly conducting side wall which we used  in all simulations ((\ref{Eq:currentout}), denoted as BC1) is not entirely consistent with the real experiment where the total current was imposed through the lateral wall. The experimental conditions would be represented by replacing (\ref{Eq:currentout})  by the electric boundary condition
\begin{equation}
\textbf{\textit{v}} =0, \quad \partial_{r}\varphi=-\frac{I}{A_{side}\sigma^{2} BU_{0}} \quad {\rm at } \quad r=r_0,
\end{equation}
denoted as BC2, and where $A_{side}$ is the surface area of the side wall.
For this reason, we conducted two further validation steps for $Ha=66$ and $Re=15972$. Firstly, we  compared the total current through the side wall with the total injected current $I$ based on BC1. The relative error of $0.3\%$ implies that the applied boundary condition is reasonable. We also compared a simulation with BC1 to one where an homogeneous current is imposed across the outer side wall (figure 3). The maximum relative errors on the mean azimuthal velocity and the rms of the fluctuations along radial direction between the two boundary conditions are less than $0.8\%$ and $3\%$, respectively. This indicates that the choice of either BC1 or BC2 does not have any significant impact on the resulting flow field and that both are compatible with good conservation of charge.

As a further validation, the numerical solutions of the average azimuthal velocity $\langle U_{\theta}\rangle_{\theta}$ and the time and space average  of the angular momentum, $\overline{L}_{lam}$, are also compared with the results produced by the Q2D model, as shown in Table.\ref{Grid}. Here, $U_{i}, (i=\theta,r,z)$ represents individual velocity components averaged over time in a quasi-steady state, $\langle \cdot\rangle_{i},(i=\theta,r,z)$ represents the average along the specific direction (Hartmann layers are excluded), and
\begin{equation}
  \langle{L}_{lam}\rangle_t=\frac{1}{T}\int^{T}_{0}  \langle L_{lam}\rangle_{V}dt, \quad \langle L_{lam}\rangle_{V}=\frac{1}{V_\Omega}\int rv_{\theta}(r)d\Omega,
  \label{Eq:averAngular}
\end{equation}
where $V_\Omega=2\pi \tilde r_0^2$ is the non-dimensional volume of the computational domain,  $\langle \cdot\rangle_{V}$ is the volume average and  $v_i$ is the instantaneous velocity.
According to the theory of \citet{Sommeria1982}, an approximate global angular momentum can be derived \citep{Potherat2000}, under the assumption of axisymmetry, \emph{i.e.}
  \begin{equation}{\label{Lameqn}}
   \overline{L}_{SM82}=\frac{I}{4\pi U_{0}a\sqrt{\rho\nu\sigma}}(1-\frac{r^{2}_{e}}{r_{0}^{2}}).
\end{equation}

 Across the range of considered parameters, the maximum relative errors of $\langle v_{\theta}\rangle_{\theta,t}/U_\theta^{\rm SM82}(r)$  and $\langle{L}_{lam}\rangle_t/{L}_{SM82}$ are less than $2.1\%$, indicating that the DNS results are reliable. Moreover, we also conducted an extensive grid
sensitivity study, the results of which are presented in the next section.
\subsection{Grid details}

Due to the localisation of the Lorentz force within $r\in[5.4, 11]$, the fluid rotates around the axis of the container, and a free shear layer forms at $r = 5.4$. In order to capture more precise flow information, highly refined grid resolution are required in both of the free shear layer and the outer wall side layer. Note that the unstructured computational grids are made of hexahedra and prisms in the present study, and the grid details in case of $Ha=792$ and $Re=15972$ are used for illustration. In the radial direction,  $N_{r}=864$ grid points are generated, 30 (\emph{resp.} 25) of which are devoted to the side wall (\emph{resp.} free shear) layer located at $r =11$  ($r=5.4$). These points are distributed within the layer according to a geometric  ratio of $\gamma_{r}= 1.1$ starting at $r = 11$ and $r = 5.4$ with an initial interval of $\Delta r_{min}\approx 0.0011$, while the
largest step (in the middle of the container) is $\Delta r_{max}\approx 0.023$.
The azimuthal direction uses $N_{\theta}=4096$ uniformly spaced grid points and the axial direction uses $N_{z}=300$ non-uniformly spaced grid points. Note that to fully resolve the Hartmann layer along the axial direction, 25 uniformly spaced grid points are devoted to each layer and a smooth transition is set toward the core region where a coarser grid resolution is sufficient. Grid points are spread  according to a geometric sequence of ratio $\gamma_{z}= 1.1$ starting at $z \approx Ha^{-1}$ and $z \approx1-Ha^{-1}$. The point nearest to the Hartmann walls is located $\Delta z_{min}\approx 5\times10^{-5}$ away from them, and the largest step is $\Delta z_{max}\approx1.2\times10^{-2}$.

In order to assess the quality of the grids, wall coordinates are introduced
\begin{equation}
   r^{+}=Re^{Sh}_{\tau}r, \quad z^{+}=Re^{Ha}_{\tau}z,
\end{equation}
where
\begin{equation}
   Re^{Sh}_{\tau}=\sqrt{Re\overline{\tau^{Sh}}}, \quad Re^{Ha}_{\tau}=\sqrt{Re\overline{\tau^{Ha}}}.
\end{equation}
\begin{equation}
   \overline{\tau^{Sh}}=\frac{1}{TA_{Sh}}\int \int (-\frac{\partial v_{\theta}}{\partial r})\Big|_{r=1}dt ds, \quad \overline{\tau^{Ha}}=\frac{1}{2TA_{Ha}}\int \int (\frac{\partial v_{\theta}}{\partial z}\Big|_{z=0}-\frac{\partial v_{\theta}}{\partial z}\Big|_{z=1})dt ds.
   \label{Eq:wallstress}
\end{equation}

Here, $\overline{\tau^{Sh}}$, $\overline{\tau^{Ha}}$ denote the associated dimensionless forms of the mean stress at the side wall and Hartmann wall, respectively. $T$ is the time interval for average, $A_{Sh}=22\pi$ and $A_{Ha}=121\pi$.
Hence, the value of the smallest wall-normal grid step in the $\Delta r^{+}$ units varies from 0.08 (see Table.\ref{Grid}). The respective variation in the $\Delta z^{+}$ units is from 0.18 (see Table \ref{Grid}). Moreover, the simulated results show that the highest velocity occurs at $r\approx 7.04$, where the azimuthal grid step $\Delta\theta\approx0.011$ is sufficiently small. Meanwhile, in order to ensure that the full range of the dissipative scales is resolved, the smallest turbulent scales ($l^{\rm min}_{\perp}$ and $l^{\rm min}_{z}$) predicted by \citet{Potherat2010} are used to evaluate the grids quality. The adopted grids indicate $\frac{\Delta^{\rm max}_{\perp}}{l^{\rm min}_{\perp}}\simeq 0.74, \frac{\Delta^{\rm max}_{\parallel}}{l^{\rm min}_{z}}\simeq 0.83$ (The smallest scales are estimated according to the turbulent Reynolds number, $Re_{t}=\frac{U_{L}S_{L}}{\nu}$). Here, $S_{L}$ is the size of the large scales, which is evaluated from the profiles of RMS of relative azimuthal velocity fluctuations (see figure 8(a) of \citet{Potherat2011}), and $U_{L}$ is the velocity of the large scales, which is calculated according to $U_{L}=(\int v_{i}^\prime v_{i}^\prime dv)^{1/2}, i=(\theta,r,z)$, where $v_{i}^\prime$ denotes the velocity fluctuations. The sizes of smallest scales according to \citet{Potherat2010} for all cases are listed in Table \ref{regime}.

\begin{table}

  \begin{center}
\def~{\hphantom{0}}
\begin{tabular}{cccccccc}
    &$N_{z}$    &$N_{Ha}$ &$N_{Sh}$    &$\overline{\tau^{Sh}}$   & $\overline{\tau^{Ha}}$  &$\langle v_{\theta}\rangle_{\theta,t}/U_\theta^{\rm SM82}(r,z)$ &$\langle{L}_{lam}\rangle_t/{L}_{SM82}$\\
   $G1$ &168 & 10  & 12 &36.3  &786.8 & 0.980 & 0.979 \\
   $G2$ &188 & 10  & 12 &36.4  &788.9 & 0.984 & 0.981 \\
   $G3$ &208 & 20  & 20 &37.3  &805.5 & 0.996 & 0.997 \\
   $G4$ &258 & 20  & 25 &37.4  &808.4 & 0.998 & 0.999 \\
   $G5$ &300 & 25  & 30 &37.5  &809.3 & 0.998 & 0.999\\
  \end{tabular}
    \caption{\label{Grid}Grid sensitivity study, $N_{Ha}$, $N_{Sh}$ denote the grid points within the Hartmann layer and side wall layer, respectively.  $\overline{\tau^{Sh}}$, $\overline{\tau^{Ha}}$ , $\langle{L}_{lam}\rangle_t/{L}_{SM82}$  and $\langle v_{\theta}\rangle_{\theta,t}/U_\theta^{SM82}(r,z) $ at $r=9.6$, $z=0.5$, which results from the different grids, are compared in case of $Ha=792, Re=15972$. }
     \end{center}

\end{table}

In addition, the grid independence studies are also conducted on the numerical case of $Ha=792$ and $Re=15972$.
Note that not only the grid sizes, but also the grid points along the $z$-direction, within the Hartmann layer and within the side wall layer are tested. The time averaged wall stress $\overline{\tau^{Sh}}$, $\overline{\tau^{Ha}}$, the time averaged angular momentum $\langle{L}_{lam}\rangle_t$ and the time-space average azimuthal velocity  $\langle v_{\theta}\rangle_{\theta,t} $, at $r=9.6$, $z=0.5$ are presented in Table \ref{Grid}.

For $Ha=792, Re=15972$, (\ref{Eq:Uteta1}) and (\ref{Lameqn}) are not strictly valid since the flow is not axisymmetric but remain sufficiently accurate to roughly assess the accuracy of the simulations.

All the simulations are stopped when the total angular momentum of the flow is statistically steady, \emph{i.e.} after $3t_{Ha}$, where
\begin{equation}
t_{Ha}=\tilde{t}_{Ha}/(a/U_{0}) \qquad{\rm and} \qquad \tilde{t}_{Ha}=a^{2}/(\nu Ha)
\label{eq:tha}
\end{equation}
 denote the non-dimensional and dimensional Hartmann damping times, respectively. Average and RMS quantities are then evaluated over a time interval of $4t_{Ha}$, and the computed values of these parameters are compared to evaluate the grid resolution within the Hartmann layer and side wall layer. Evidently, the results are very close to each other, even with the worst spatial resolution, as shown in Table.\ref{Grid}. Firstly, the reliability and the accuracy of the DNS results are confirmed by a difference of less than $2.1\%$ between the present results and the solutions predicted by (\ref{Eq:Uteta1}) and (\ref{Lameqn}).
Moreover, grid-independent solutions are also achieved with the grids under consideration. For example, less than $3.2\%$ difference is found for all the predicted values on the coarsest grid and the finest grid, and this discrepancy is even further reduced  when the two finest grids are compared. Hence, one can conclude that grid $G5$ is sufficiently fine to simulate the flow in the case of $Ha=792$ and $Re=15972$. The mesh is, however, further refined when 3D effects become significant, e.g. in the case of $Ha=55$ and $Re=15972$.

For different numerical cases, the dimensionless time steps used in the computations are altered according to the parameters applied in simulation, such that $\Delta t=5.0\times10^{-5}$ for $(Ha,Re)=(264,4791)$,
$5.5\times 10^{-5}$ for $(Ha,Re)=(264,31944)$ and $1.6\times 10^{-5}$  for all other cases. The values are determined by the limits
of numerical stability, which highly depends on the viscous term  and  the convective term of the momentum equations at high $Re$. In addition, higher $Re$ or $Ha$ demand smaller time steps because of the higher azimuthal velocity and the thinner Hartmann layers.

\section{Results and discussion}\label{sec:3}

\subsection{Validation of velocity profile at $Ha\gg1$ and $N\gg1$}\label{sec:3.0}

  \begin{figure}
  \centering
  
  \includegraphics[width=0.48\textwidth]{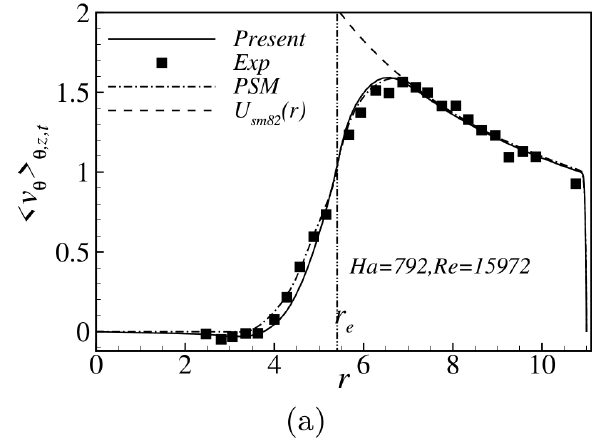}
    \includegraphics[width=0.48\textwidth]{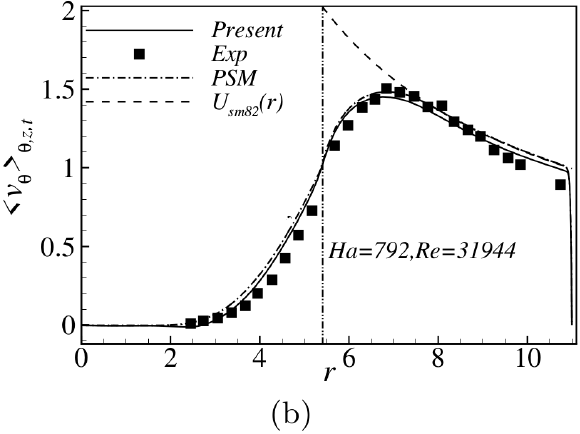}
  \caption{\label{r_profile} Comparison of $\langle v_{\theta}\rangle _{\theta,z} $  between experiment  (symbols), present numeric (solid lines), 2D numeric based on PSM model(dashed dot lines) and $U_\theta^{\rm SM82}(r)$ (dashed lines) for cases at $Ha=792$, $Re=15972$ (\textit{a}) , and $Ha=792$, $Re=31944$ (\textit{b}). The dashed line depicts an algebraic law $r^{-1}$  (predicted by Eqn.(\ref{Eq:Uteta1})).}
\end{figure}

As far as the authors know, it is the very first attempt to reproduce the MATUR experiment by performing 3D direct numerical simulations, and hence as a necessary validation procedure, a detailed comparison with the available experimental results need to be carried out. In addition, since the PSM model can deal with the cases when $Ha\gg 1$ and  $N_{t}=N(\widetilde{r_{0}}/a)^{2} \gg 1$, some numerical cases falling into this space are also investigated for validation. However, note that this model becomes imprecise when either of the parameters, $Ha$  or $N_{v}$ become of the order of 1, ($N_{v}=N(\widetilde{r_{0}}/a)$ is the interaction parameter based on the horizontal scale), which, again, stresses the importance of conducting 3D direct numerical simulations.

\begin{figure}
  \centering
  
  \includegraphics[width=0.46\textwidth]{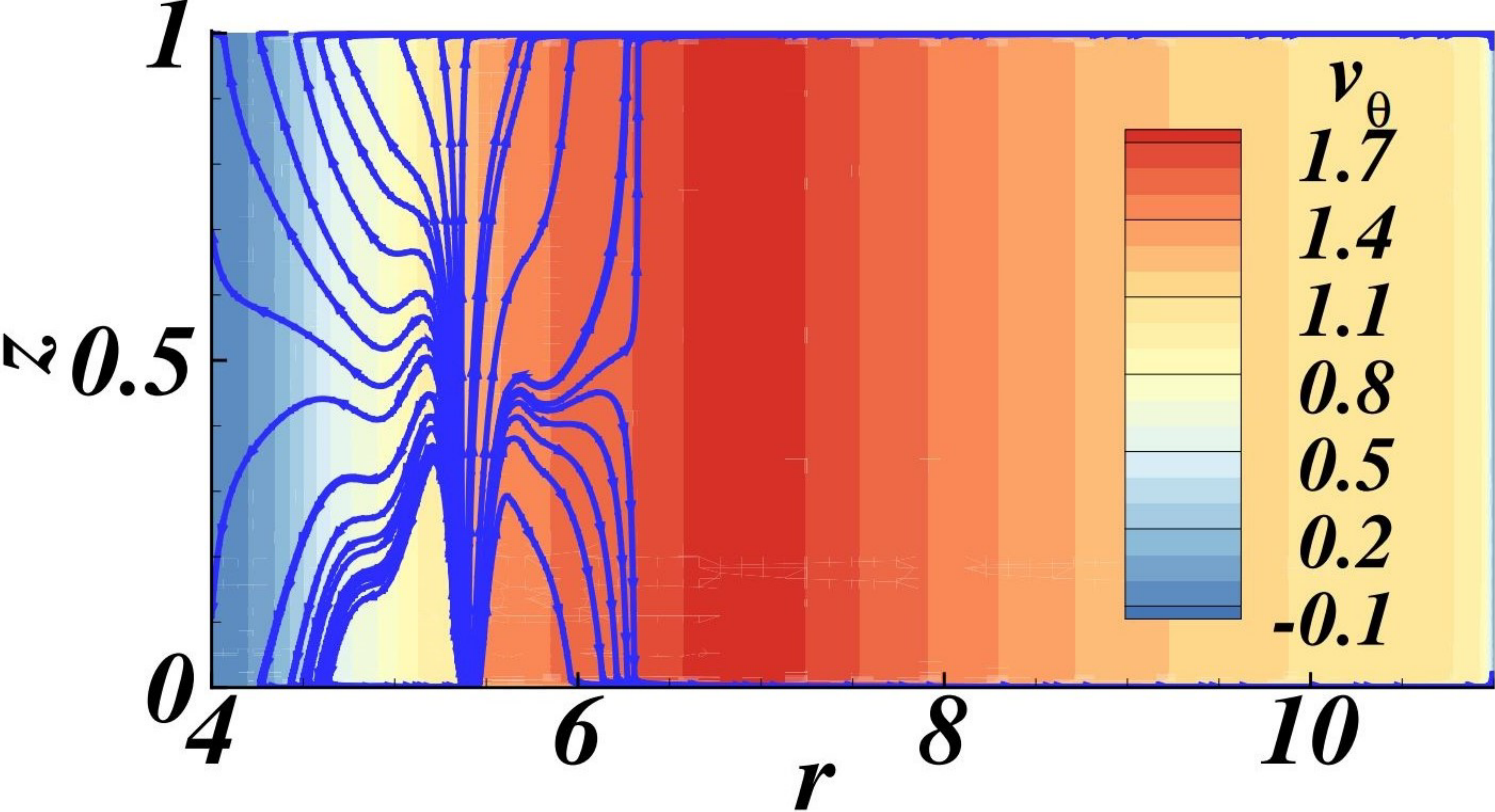}\\
    \includegraphics[width=0.45\textwidth]{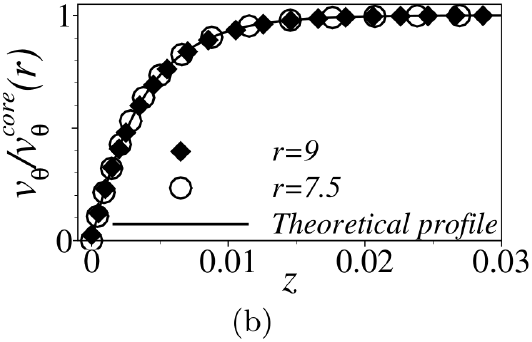}
      \includegraphics[width=0.45\textwidth]{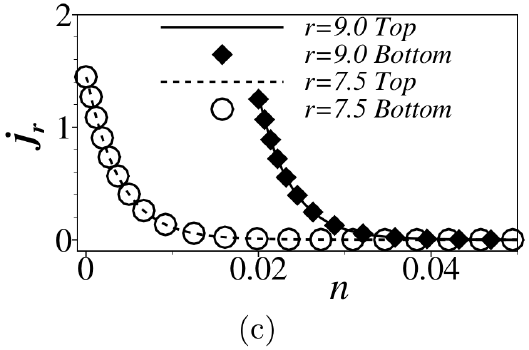}
  \caption{\label{ucyl}(\textit{a}) The distribution of the instantaneous azimuthal velocity and current streamlines on plane $\theta=0$ at $Ha=264$, $Re=15972$. (\textit{b}) The vertical profiles of the instantaneous azimuthal velocity  within the bottom Hartmann layer along $r=7.5$ and $r=9$. (\textit{c}) the distribution of radial current density both across top Hartmann layer and bottom Hartmann layer along $r=7.5$ and $r=9$, where the origin of the distribution along $r=9$ is shifted to $n=0.02$, where $n$ is the wall normal coordinate \emph{i.e.} for the bottom wall, $n=z$, and for the top wall $n=1-z$ and $v_\theta^{core}(r)=\langle v_\theta(r,\theta,z=0.03,t)\langle_{\theta,t}$. Here $v_\theta$ is averaged in time and $\theta$.}
\end{figure}

The predictions of the mean azimuthal velocities from different approaches, denoted by $\langle v_{\theta}\rangle _{\theta,z} $, are plotted in figure \ref{r_profile}.
Clearly, a good agreement is found between the experimental results, the numerical data and the laminar theoretical prediction. In the outer region $r_{e}<r<r_{0}$, the maximum relative discrepancy  between the results of DNS and experiment  is less than $6.7\%$ $(8.5\%)$ in the case of $Re=15972$ $(Re=31944)$ based on an error estimation of $\frac{\|\langle v_{\theta}^{\rm DNS}\rangle _{t}-\langle v_{\theta}^{\rm exp}\rangle _{t}\|_{2}}{\|\langle v_{\theta}^{\rm exp}\rangle _{t}\|_{2}}$. The maximum relative discrepancy between the results of DNS and the PSM model is less than $1.0\%$ $(2.9\%)$ in the case of $Re=15972$ $(Re=31944)$ based on an error estimation of $\frac{\|\langle v_{\theta}^{\rm DNS}\rangle _{t}-\langle v_{\theta}^{\rm PSM}\rangle _{t}\|_{2}}{\|\langle v_{\theta}^{\rm PSM}\rangle _{t}\|_{2}}$. In particular, the velocities exhibit the characteristic feature that they increase sharply across the free shear layer ($r=r_{e}$) due to the current injection. Accordingly, the shear layer separates the flow into an outer and inner regions. Moreover, from both of the experimental and the numerical data, the downward trend of the azimuthal velocity  in regions between the injected electrodes and the vertical wall follow the expected scaling law of $\langle v_{\theta}\rangle _{\theta,z}\sim r^{-1}$, as predicted by (\ref{Eq:Uteta1}), and which reflects the geometrical spreading of the radial forcing current in the Hartmann layer, i.e. $ j_{r}\sim (4\pi r)^{-1}$ (see figure \ref{ucyl}(c)).

A typical instantaneous distribution of $v_\theta$ obtained numerically over a radial cross-section $\theta=0$ for moderate forcing current is shown in figure \ref{ucyl}(a).
Under a strong magnetic field, the velocity gradient along the magnetic field lines is remarkably damped, except in the Hartmann layers
where an exponential profile subsists. The detailed velocity distribution in the Hartmann layer is shown in figure \ref{ucyl}(b), and a good agreement is observed between the present numerical results and the exact solution, given as $v_{\theta}=v_{\theta}^{\rm core}(1-$exp$(Haz))$, with $v_{\theta}^{\rm core}$ indicating the azimuthal velocity in the core flow. It also demonstrates that the thickness of the Hartmann layer at $r=7.5$  and $r=9$ is the same. Besides, figure \ref{ucyl}(c) reveals that the vertical profiles of radial current density within the top and bottom Hartmann layers collapse with each other, implying that the electric current intensity $I$ injected at the electrodes divides in two equal parts between the two symmetric Hartmann layers. In addition, the the radial current density is much higher 
near the Hartmann wall, so the Joule dissipation mainly takes place in the thin Hartmann layers, in line with the laminar Hartmann layer theory.
Therefore, the annular fluid domain located between the selected circular electrodes and the cathode $(5.4\leq r \leq 11)$ is driven in the azimuthal direction by the Lorentz
force, while the central fluid domain $(r <5.4)$ is entrained by friction within the free shear layer.
Interestingly, the current density at the upper wall stands a little lower than at the bottom wall, showing that despite the excellent agreement between PSM and experimental data, the flow is ever so slightly three-dimensional.

In the following part, the evolution of the large structures and the spectral analysis are discussed. Subsequently, we study the secondary flow induced by Ekman pumping. We also investigate the characteristics of the free shear layer and the side wall layer before presenting the turbulent statistics, global angular momentum and three-dimensionality.

\subsection{General behaviour of the flow}\label{sec:3.1}

\begin{table}
  \begin{center}
\def~{\hphantom{0}}
\begin{tabular}{ccccccccccccc}
   $Ha$                             &264   &528    &792    &792   &264   &264    &132    &110     &99     &80    &66     &55 \\
   $Re$                             &4792  &15972  &15972  &31944 &15972 &31944  &15972   &15972  &15972  &15972 &15972  &15972 \\
$N_{v}$                             &158.9 &190.6 &428.9   &214.5 &47.7  &23.9   &11.9    &8.3     &6.7   &4.4   &3.0    &2.1 \\
   $R$                             &18.2   &30.3  &20.2    &40.4  &60.5  &121.0  &121.0   &145.2  &161.3  &199.6 &242.0  &290.4\\
$l^{min}_{\perp} (\times 10^{-2})$ &3.95   &3.05  &3.10    &2.52  &2.91  &2.45   &2.88    &2.84   &2.81   &2.77   &2.71   &2.67 \\
$l^{min}_{z} (\times 10^{-2})$     &1.92   &1.31  &1.45    &1.34  &1.22  &1.12   &1.10   &1.08   &1.01    &0.96   &0.92   &0.89\\
  \end{tabular}
    \caption{\label{regime}Non-dimensional parameters in cases calculated numerically.}
     \end{center}
   \end{table}

The different cases investigated are listed in Table \ref{regime}, where the dimensionless parameter $R(=Re/Ha)$ represents the Reynolds number scaled on the thickness of the Hartmann layer. According to the experiments of \cite{Moresco2004}, the flow within the Hartmann layer becomes turbulent when $R \geq 380$, in which case the DNS will require enormous computational resources. Therefore, only cases with  $R<380$ are considered in this paper. Furthermore, five of the relevant interaction parameters scaled on the horizontal length  $N_{v}$ are at the order of unity, aiming to study the three dimensionality of the flows.

\begin{figure}
  \centering
   \includegraphics[width=0.9\textwidth]{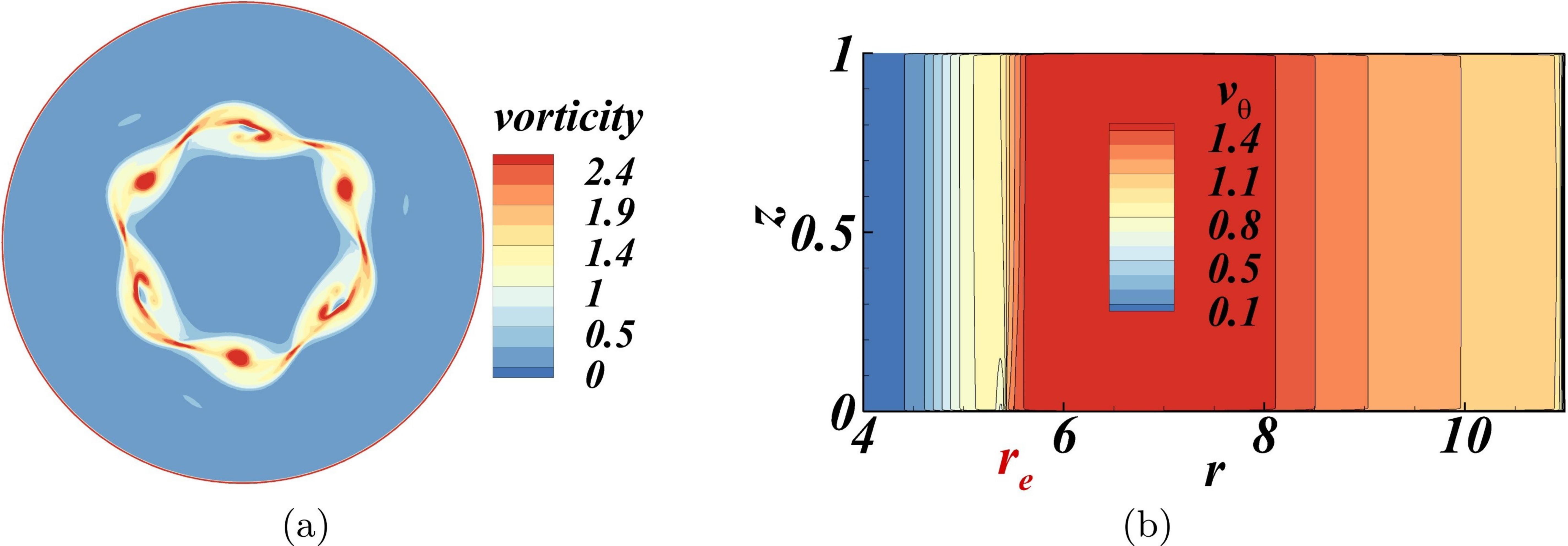}
   \includegraphics[width=0.9\textwidth]{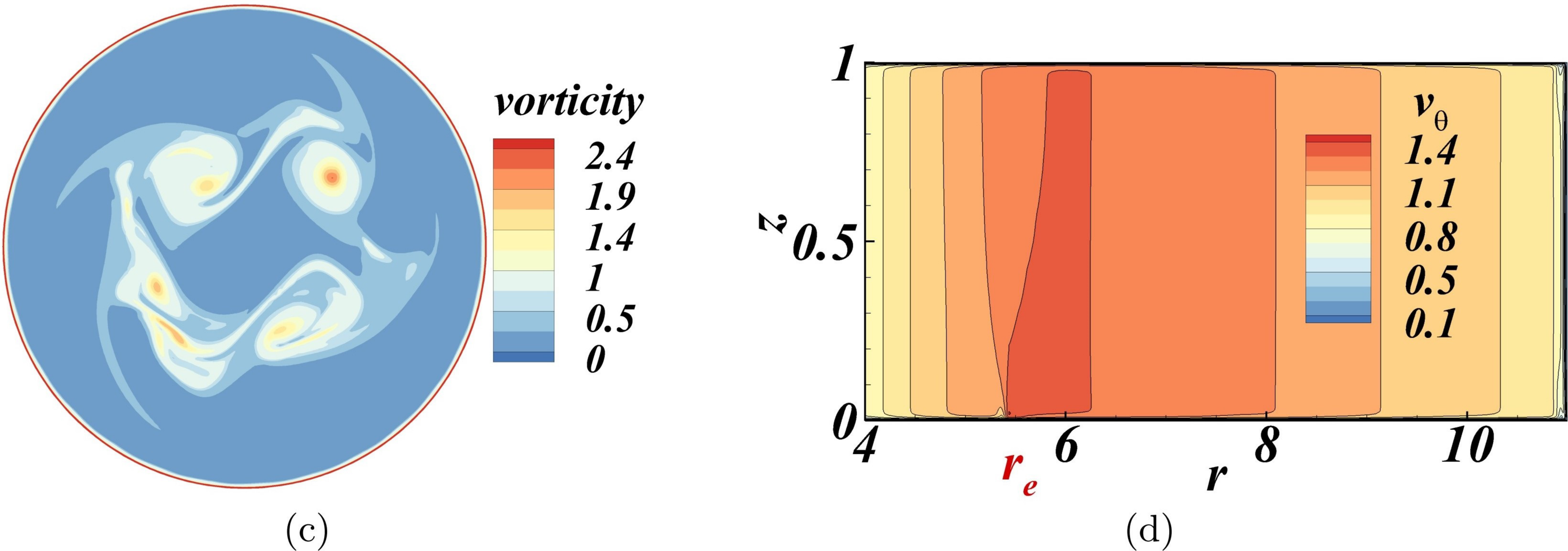}
   \includegraphics[width=0.9\textwidth]{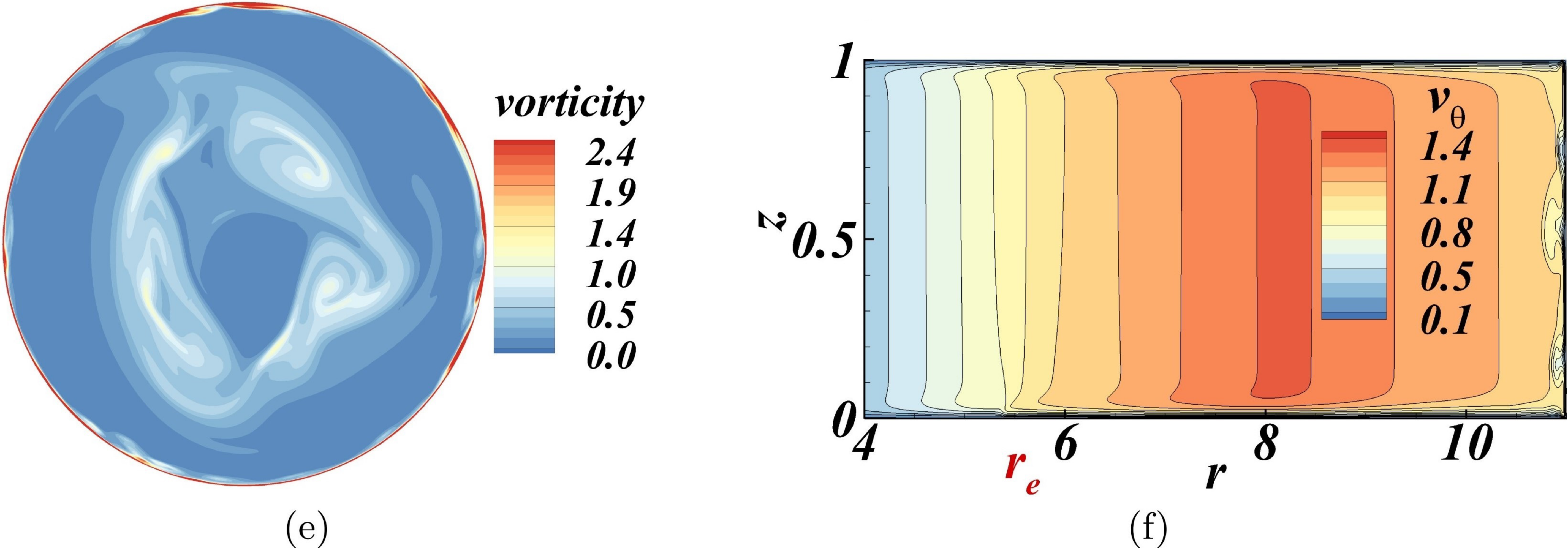}
  \caption{\label{regimePic}Typical snapshots of equilibrium or quasi-equilibrium states obtained from numerical
simulations. Contours of the magnitude of the vorticity on plane $z=0.5$ (left column), contours of instantaneous azimuthal velocity on plane $\theta=0$ (right column). (\textit{a}), (\textit{b})  $Ha=792, Re=15972$.  (\textit{c}), (\textit{d})  $Ha=264, Re=31944$. (\textit{e}), (\textit{f})  $Ha=55, Re=15972$. The velocity are normalised  with $U_{0}$.} 
\end{figure}

In the calculations, the electrical current is injected at $t=0$ when the fluid is at rest and remains constant during the whole simulation, following the actual experimental procedure. For different electrical current intensities, the flow goes through a sequence of evolution and reaches different equilibrium and quasi-equilibrium states  presented on figure \ref{regimePic}. We shall now give an overall view of the numerical results, while more details of the evolution  and local quantities will be reported later.

For $R\leq 121$, the evolution of the flow is qualitatively similar to that found for $249\leq R\leq 1122$ in two-dimensional simulations of MATUR \citep{Potherat2011}. In this work, the current was injected at the same location as in the present work, and there was no separation of the side layer, but the Hartmann layer was modelled as turbulent for $R\geq380$.
The flow contains five or six relatively stable vortices rotating around the $z$ axis in near-solid body
rotation.
They mainly remain localized near the free shear layer once generated there, as shown in figure \ref{regimePic}(a). Thus, the velocity fluctuations in the inner region and the annular outer region are of much lower intensity, and the azimuthal velocity contours reveal that the wall side layer and the Hartmann layer are stable. Besides the low value of $R$, part of the reason for the stability of the side layer is that these large vorticity structures remain distant from it, and little interaction between them takes place. However, the thickness of the side wall layer is still  smaller than the scaling for a straight duct ($Ha^{-1/2}$), due to the recirculations induced by Ekman pumping, a point we will
analyse in detail in $\S$ \ref{sec:3.3}.
Since Most of the large vortices remain near the centre of the domain, highly turbulent fluctuations are induced there. By contrast, the velocity fluctuations in the annular region are much weaker, especially near the side wall. It is the tail of the vortices that causes the velocity fluctuations there, as it is  stretched and conveyed outwards. The induced flow in the outer region therefore exhibits long azimuthal vorticity streaks and much lower fluctuation intensity than in the inner region.

For $R\geq121$ small scale three-dimensional turbulence appears in the side layer.
The onset of three-dimensional turbulence within $ R\lesssim 121$ is consistent with the value of 138 reported by \cite{zhao2012}, albeit a little lower. The difference in curvature of the external wall ($a/\tilde r_0=1/9$ in MATUR and $a/\tilde r_0=1/5$ (in our notations) in \cite{zhao2012}, suggests that recirculations may be more important in the latter than the former. Since their effect is rather stabilising, this could explain the lower value detected here.
%
%

\begin{figure}
  \centering
  \includegraphics[width=0.9\textwidth]{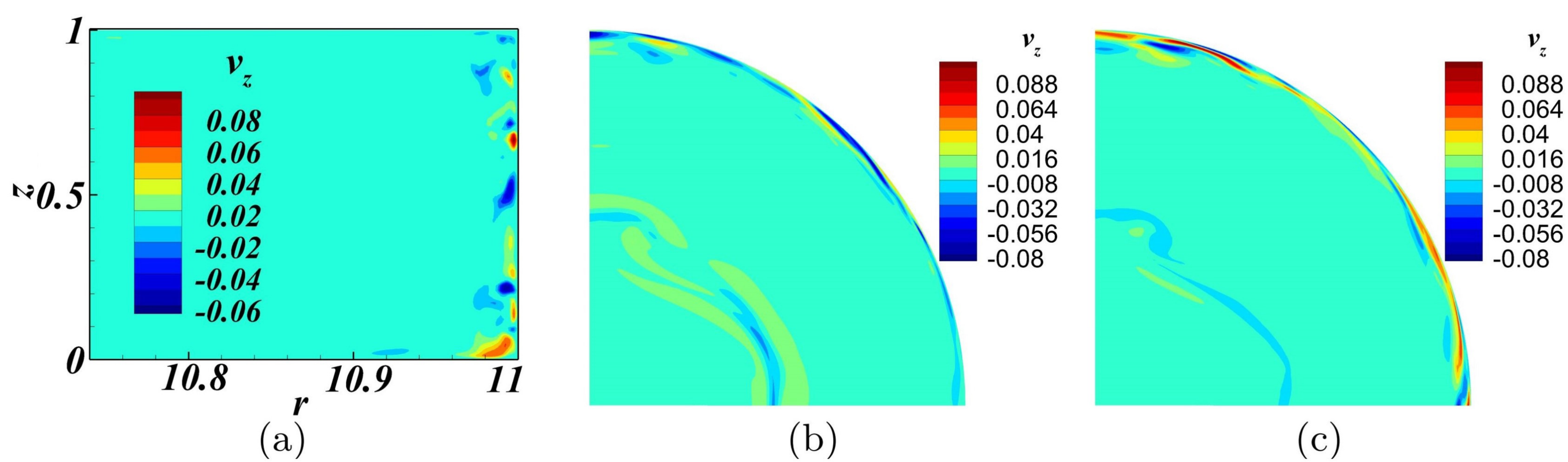}
  \caption{\label{structure}  Snapshots of velocity contours for $Ha = 55$, $Re = 15972$. (\textit{a}) Distributions of axial velocity $v_z$in the plane $\theta = 0$ (near the side wall). 
(\textit{b})  Distribution of vorticity in the plane near the top Hartmann wall at $z=0.8$.  (\textit{c})  Distribution of vorticity in the plane near the bottom Hartmann wall at $z=0.2$. Note that the contour levels for vorticity and $v_z$ are chosen so as to enhance the visibility of turbulent flow structures.}
\end{figure}

For $R\geq145.2$, the size of the large structures increases, leading to highly turbulent fluctuations in
 both the inner and the outer regions. Accordingly, long azimuthal vorticity streaks of relatively high intensity exist in the outer region that induce instabilities within the wall side layer. Concurrently, the wall side boundary layer separates from the wall, suggesting that separation
results from the fluctuations in the outer region, as observed previously observed by \cite{Potherat2011}.
For $Ha=55, Re=15972$ ($R=290$), Ekman recirculations are strong, the Hartmann layer is relatively thick so the centripetal radial velocity in the Hartmann layer is relatively strong (see figure \ref{meanEkman}). Thus, in the vicinity of a free shear layer, figure \ref{regimePic}(f) conveys that the  azimuthal velocity  near the Hartmann layer is higher than in the core.  Within the side wall layer, one can observe the dramatic variation of the azimuthal velocity and vertical velocity (see figure \ref{structure}(a)) along the magnetic field lines besides the separation of the layer.

Finally, in the examples shown here, three dimensional effects are only noticeable within the shear layer for $R\geq121$ ($Ha$=132, $Re=15972$, this is in fact better seen from the analysis of the vertical velocity in the side layer in section \ref{sec:3.3}). 
By contrast, for $R\simeq290$ ($Ha=55$, $Re=15972$), weak three-dimensionality, where flow patterns are topologically identical but less intense near the top wall, exists outside the shear layer (see figure \ref{structure}(b), (c)). The presence of three-dimensionality, however, is controlled by the true interaction parameter at the scale of the considered structure.
A consequence is that inertia-induced three-dimensionality is expected to appear in a parallel layers of thickness $\delta_{||}\sim a Ha^{-1/2}$ when the local turnover time $\delta_{||}/U_0$ becomes smaller than the two-dimensionalisation time at that scale $\rho a^2/\sigma B^2 \delta_{||}^2$, \emph{i.e.} when the Reynolds number based on the parallel layer thickness $R_{||}=U\delta_{||}/\nu$ exceeds unity.
By contrast, since the separation of the wall-side layer is induced by the tail of large two-dimensional structures, which is mostly quasi-two dimensional, it can be expected to be controlled by $R$.
\subsection{Detailed evolution of the flow}\label{sec:3.2.}
\begin{figure}
  \centering
  
  \includegraphics[width=0.45\textwidth]{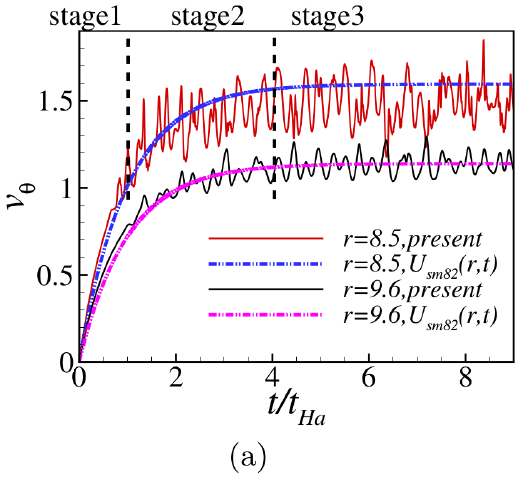}
    \includegraphics[width=0.45\textwidth]{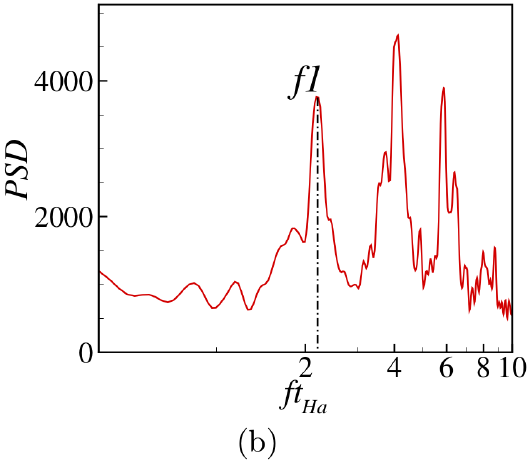}
  \caption{\label{Ucompare}(\textit{a})Typical instantaneous azimuthal velocity signals \emph{vs.} time at $(6.85,0,0.5)$ and $(9.6,0,0.5)$ at $Ha=132, Re=15972$. The solid lines denote the numerical results and the dashed lines denote the theoretical value derived by \citet{Messadek2002}, i.e.  $U_\theta^{\rm SM82}(r,t)=\frac{I}{4\pi rU_{0} \sqrt{\sigma\rho\nu}}(1-$exp$(-\frac{t}{t_{Ha}}))$. (\textit{b}) Corresponding power density spectra at point $(5.4,0,0.5)$ within the free shear layer.}
  \end{figure}

 For flows without boundary layer separation, the parameters $Ha=132$, $Re=15972$ are chosen to illustrate the typical evolution process. According to the variation of the azimuthal velocity signals, as shown in figure \ref{Ucompare}(a), three different stages can be distinguished before the flow reaches its final quasi-equilibrium state. The acceleration of the fluid in a laminar regime corresponds to stage 1.
After a short time, the laminar shear layer at $r =5.4$ becomes visible as the external annular region $5.4 \leq r \leq 11$ is driven in rotation by the Lorentz force.

 \begin{figure}
  \centering
  \includegraphics[width=0.9\textwidth]{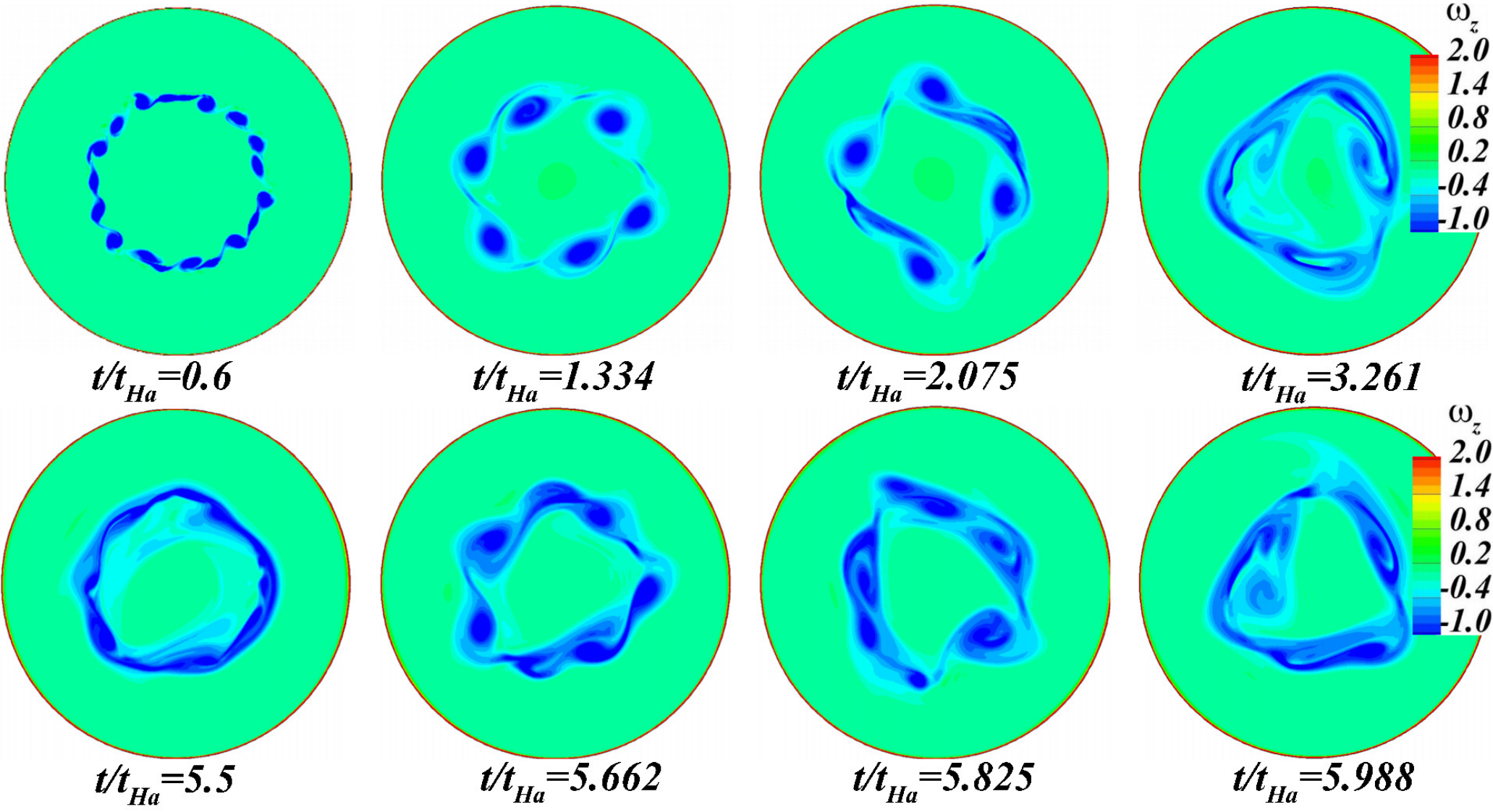}
  \caption{\label{vortvstime} Evolution of the flow with time at $Ha=132$, $Re=15972$. The distribution of the axial vortex structures, $\omega_{z}$, on plane $z=0.5$.} 
\end{figure}

\citet{Potherat2005} estimated the stability threshold for this layer as $Re/\sqrt{Ha}<2.5$, implying that the circular free shear layer becomes linearly unstable when the Reynolds number based on its thickness exceeds the threshold of 2.5. For all injected current intensities considered here, this critical value on the azimuthal velocity is reached very quickly. Then the circular free shear is subject to a Kelvin-Helmholtz instability, which breaks it up into small vortices ($t/t_{Ha}=0.6$, figure \ref{vortvstime}(a)). This process defines stage 2.
The detailed evolution of the vortices along the axial direction, $\omega_{z}$, is presented in figure \ref{vortvstime}. These vortices merge into larger structures very soon after their inception ($t/t_{Ha}=1.334, ~ 2.075, ~ 3.261$ \ref{vortvstime}(b-d)). They become distorted because of the shear. As shown in figure \ref{Ucompare}(a), the velocity at two representative locations keeps increasing during this stage. The amplitude of velocity oscillation at $r=8.5$ being larger than that at $r=9.6$, the turbulent structures exert a weaker influence in the region far from the electrodes.

Stage 3 corresponds to the quasi-equilibrium state of the flow, which, at this point, cycles through a recurring sequence. First, large structures progressively loose intensity and elongate along the free layer ($t/t_{Ha}=5.5$). Second, these segments break up again and give birth to several smaller vortex structures ($t/t_{Ha}=5.662$), which interact with each other and merge into bigger structures ($t/t_{Ha}=5.825$). Subsequently, a small number of large structures are formed ($t/t_{Ha}=5.988$). The cycle of this recurring sequence is consistent with the base frequency of velocity signals ($f1=2.26$) shown in figure \ref{Ucompare}(b).

 \begin{figure}
  \centering
  \includegraphics[width=0.9\textwidth]{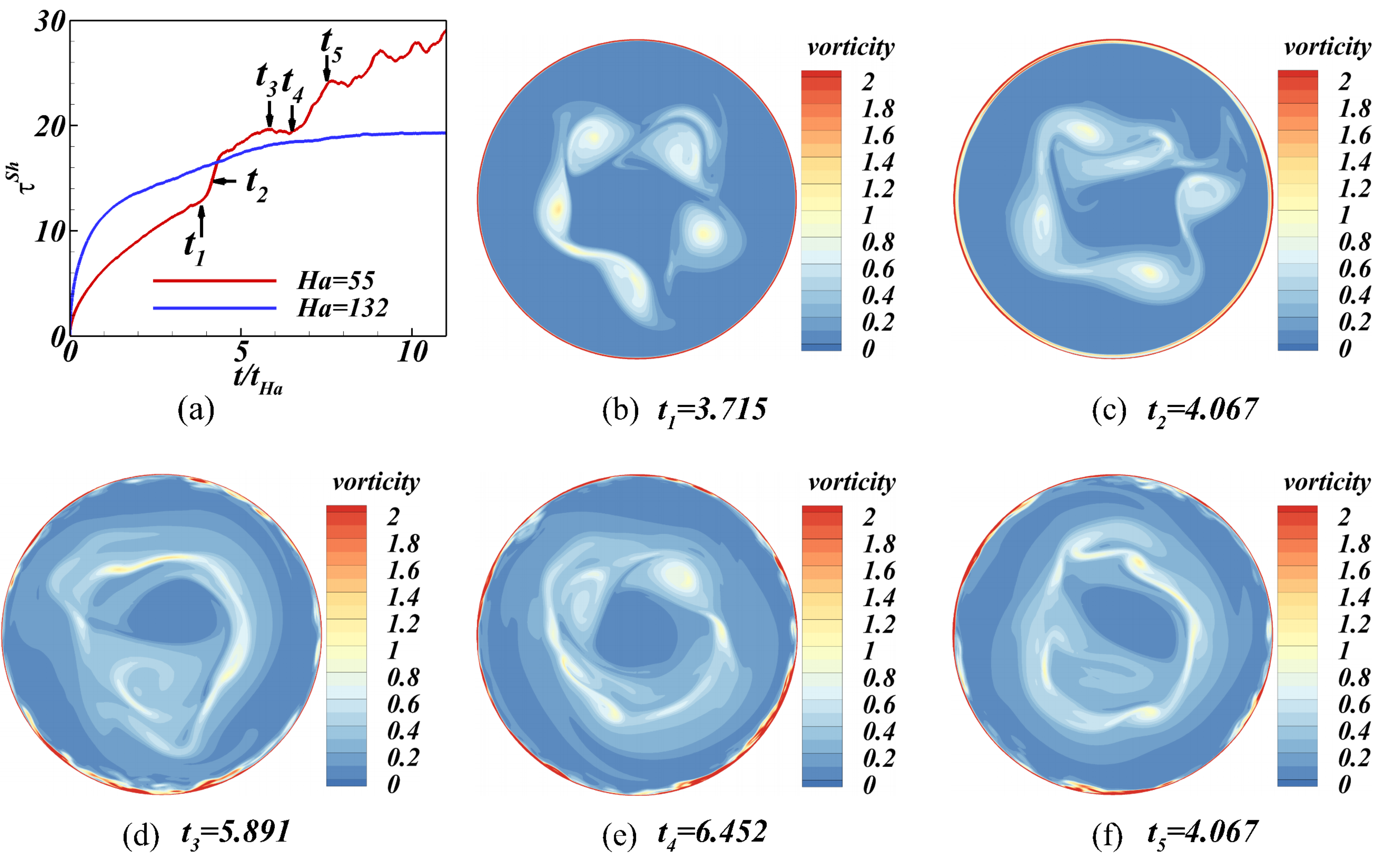}
  \caption{\label{tauRegime3}(\textit{a})The average side wall shear stress signals variation with time in case of $Ha=55$, $Re=15972$  and  $Ha=132$, $Re=15972$. (\textit{b})-(\textit{f}) Contours of the magnitude of the vorticity on plane $z=0.5$ at different time at $Ha=55$, $Re=15972$.}
\end{figure}

As $R$ is increased, the most striking feature is the appearance of separation and turbulence within the side wall layer.
Figure \ref{tauRegime3}(a) shows the evolution of the mean shear stress on the side wall $\tau^{Sh}$, with a sudden increase of $\tau^{Sh}$ between $t_{1}$ and $t_{2}$ for $Ha=55, Re=15972$. In this interval, the increase of $\tau^{Sh}$ can be ascribed to the random turbulent fluctuations, seen more in detail in section \ref{sec:3.3}.
From $t_2$ (figure \ref{tauRegime3}(c)), vorticity streaks attached to the vortices generated at the free shear layer start reaching out to the outer side layer, and incur local variations of its thickness. At $t_3$ (figure \ref{tauRegime3}(d)), these variations have become severe to the point of incurring boundary layer separation. 
The decrease of  $\tau^{Sh}$ at $t_{4}$ shown in figure \ref{tauRegime3}(a), confirms the occurrence of separation at the wall side layer. This is consistent with the visualizations of the vortex structures shown in figure
 \ref{tauRegime3}(d). For $Ha=132$ and $Re=15972$, by contrast, the evolution of $\tau^{Sh}$ is rather smooth and no brutal change in shear stress is observed, indicating the absence of separation. Moreover, the evolution of case at $Ha=55$, $Re=15972$ in the later stage ($t_{3}$, $t_{4}$, $t_{5}$) is similar to that of case at $Ha=132$, $Re=15972$, which goes through the same recurring sequence.

\subsection{Spectral analysis}\label{sec:3.2}

  \begin{figure}
  \centering

\includegraphics[width=0.42\textwidth]{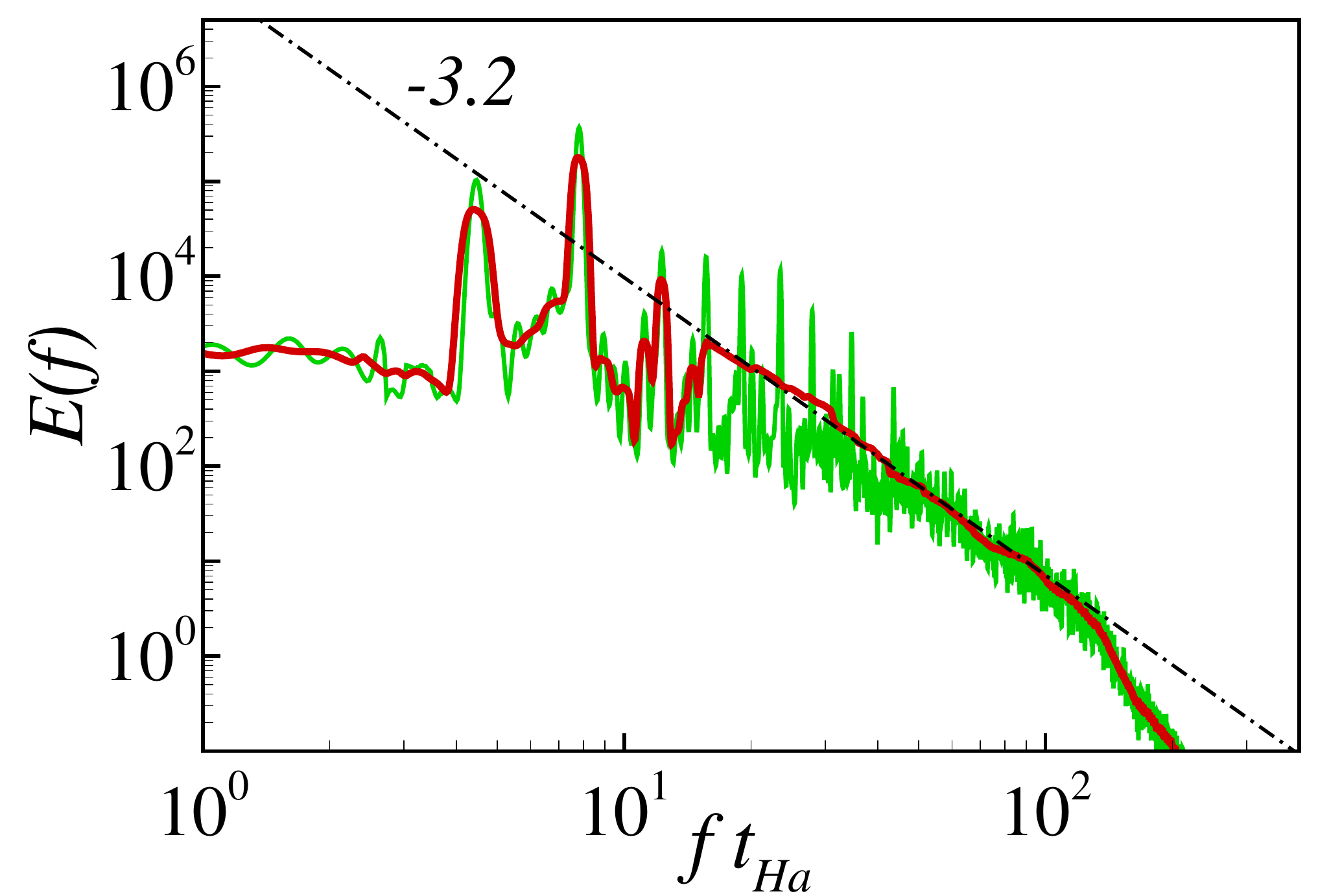}
\includegraphics[width=0.42\textwidth]{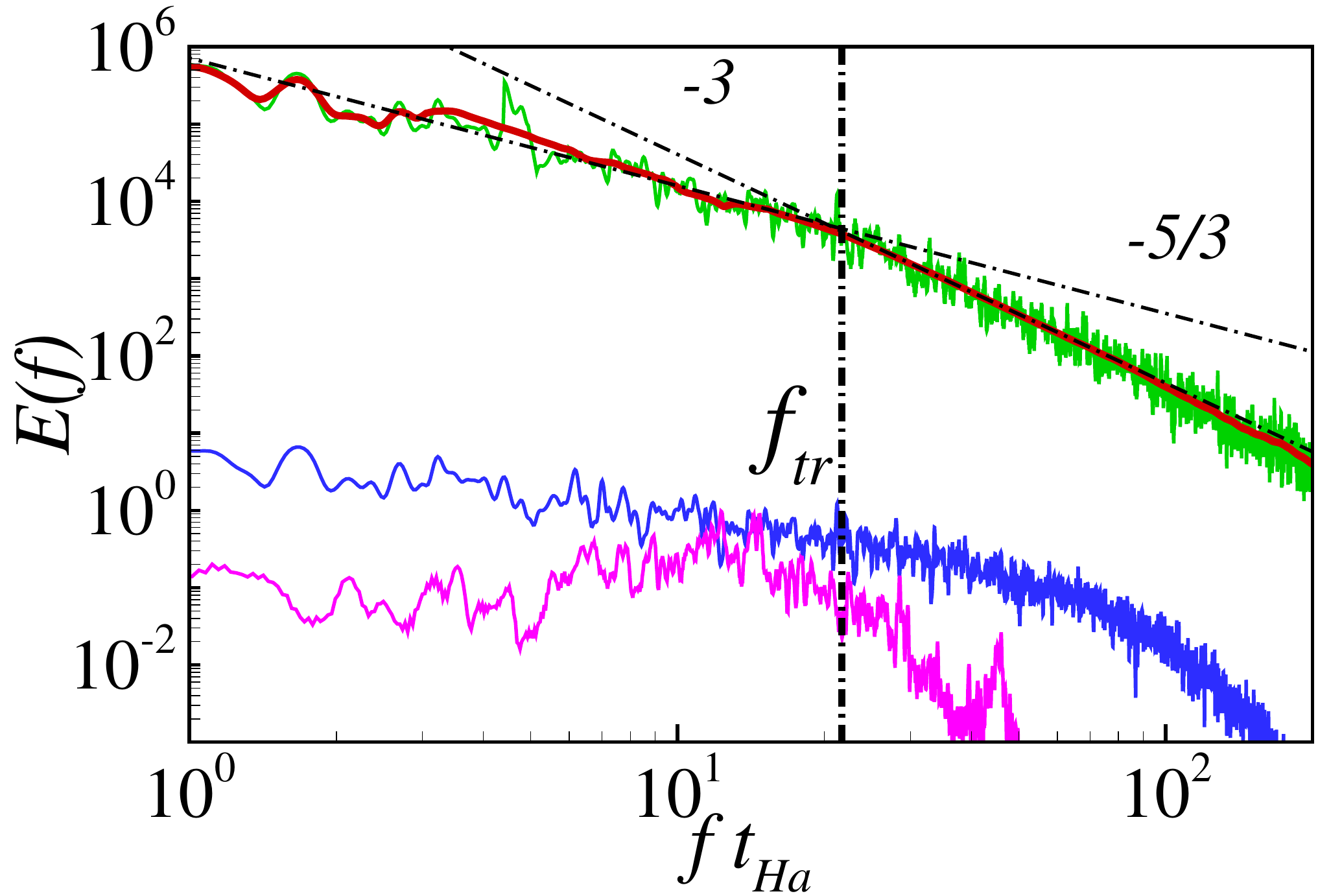}
\includegraphics[width=0.42\textwidth]{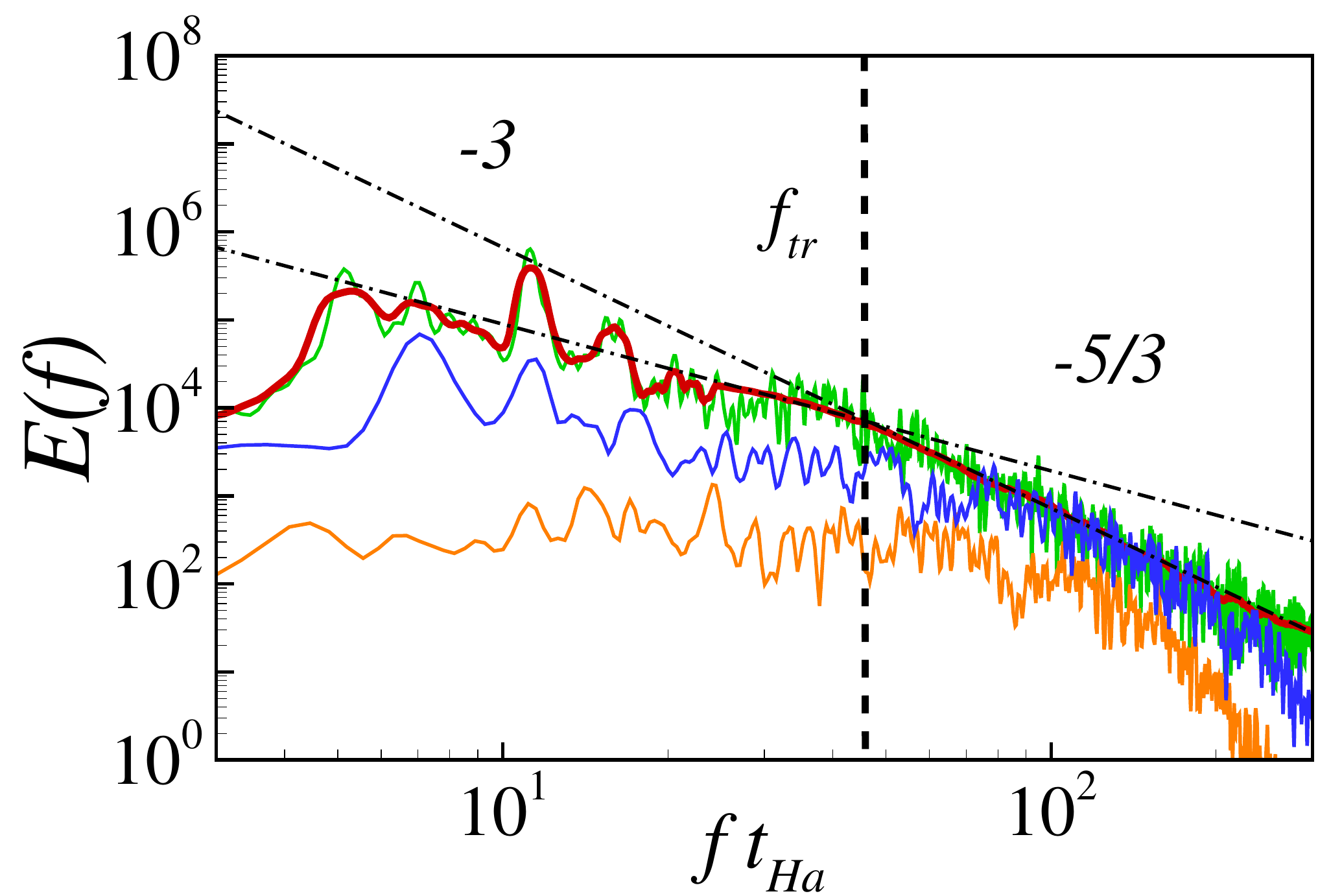}
\includegraphics[width=0.42\textwidth]{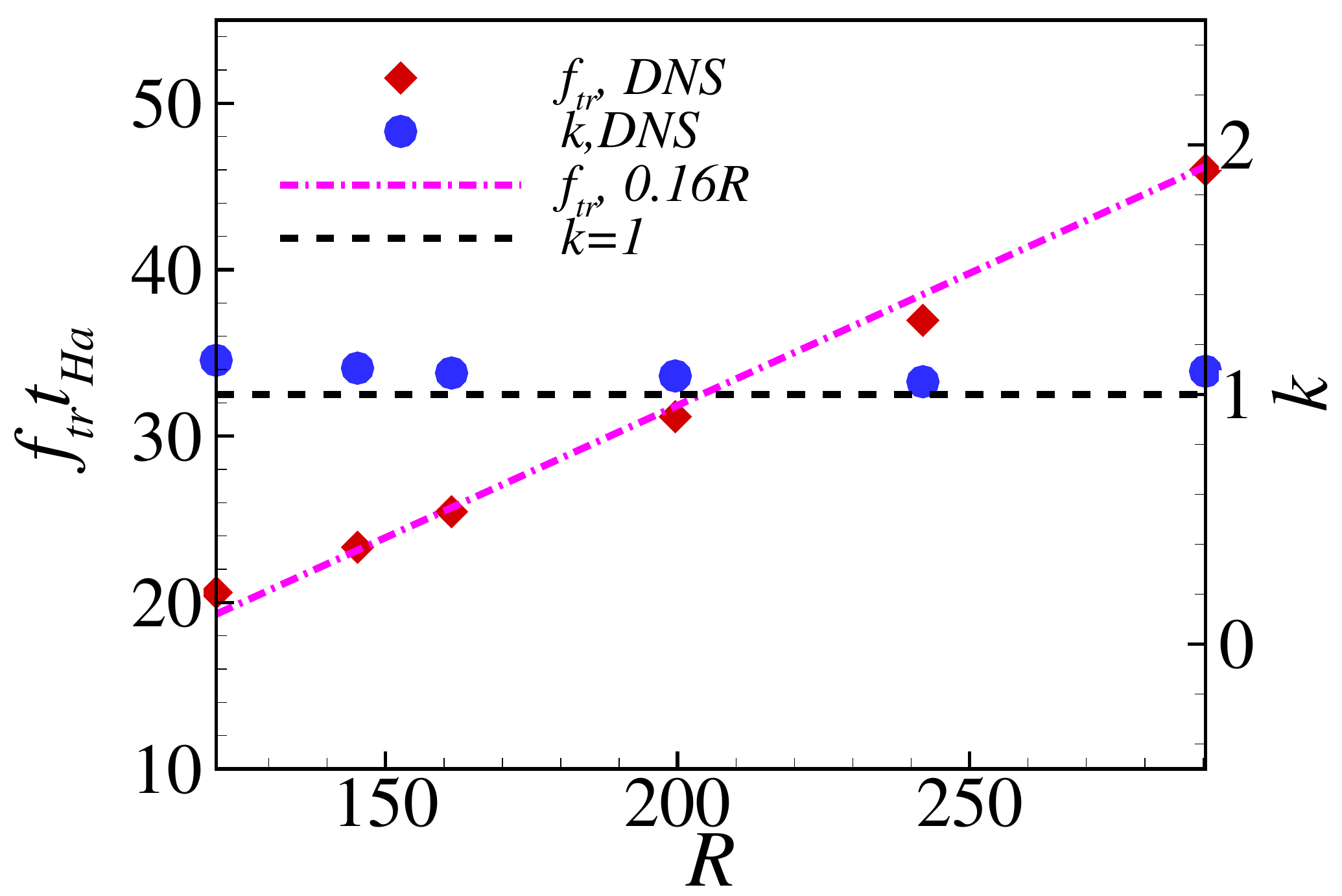}
\caption{\label{f123} Frequency spectra obtained from  velocity time-series. For reference,
the power laws $f^{-3.2}$, $f^{-5/3}$ and $f^{-3}$, denoted with dash-dot line, are also shown. Solid lines correspond to 
spectra  filtered using Bezier spline (red), time-series of  
azimuthal velocity time-series at point ($r=6.85$, $z=0.5$, green (within the free shear layer)), 
axial velocity  at point ($r=6.85$, $z=0.5$, red), 
axial velocity  at point ($r=$10.98 (b) 10.95 (c) , z=0.5, blue (within the side wall layer)).   
azimuthal velocity  at point ($r=$10.95, $z=0.5$), 
(\textit{a}) $Ha=528$, $Re=15972$ ($R=30.3$)  
(\textit{b}) $Ha=264$, $Re=31944$ ($R=121$), 
(\textit{c}) $Ha=55$, $Re=15972$ ($R=290.4$), 
(\textit{d}) The separation frequency  of the spectra slops $f^{-5/3}$ and $f^{-3}$ and the forcing scales for different $R$.} 
\end{figure}

The power density spectra of several typical cases are analysed in this section shown in figure \ref{f123}.
To calculate of the average, we employ 60 probes, uniformly distributed along the angular direction, and take the average of the measured signal as $v(t)=\frac{1}{60}\Sigma v_{i}(t)$. Spectra are obtained using Welch's averaged periodogram method, while a Hamming window was applied to each overlapping segment of data. Additionally, spatial energy spectra can be deducted from these spectra by taking advantage of the large average azimuthal velocity and using Taylor’s hypothesis, $2\pi f=\langle U_{\theta}\rangle_{\theta} k$.

For $R=30.3$, the spectrum of $v_{\theta}(t)$  exhibits an expected strong peak at a fundamental frequency corresponding to the passage of the large structures through the measuring probes. The other noticeable peaks represent the harmonic and sub-harmonic frequencies, as shown  in figure \ref{f123}(a). The scaling of the  spectrum in the inertial region obeys scales as $f^{-3.2}$. According to \citet{Eckert2001}, the spectral exponent in duct flow turbulence for $N\approx 39.0$ relevant to this case is about -3.5, which is consistent with our findings, despite the difference in the geometries of these two types of turbulent shear flows.

When $R=121$, the peak of the base frequency in the
spectra of azimuthal velocity fluctuations can still be observed (see figure \ref{f123}(b)), and thus the large structures continue to rotate around the axis. The spectrum in the inertial region  can be separated into two parts, with a transition frequency 
$f_{tr}t_{Ha}\simeq 20$. For $f_{tr}t_{Ha}\leq 20 $, the power spectral density scales as $f^{-5/3}$, while $f_{tr}t_{Ha}> 20 $, 
is scales as $f^{-3}$. By applying the Taylor's hypothesis, we find the corresponding azimuthal wavenumber of
$k_{tr}\simeq 1.10$.
This value is in accordance with the results of \citet{Messadek2002}, i.e. $\tilde k_{tr}\simeq 1 $ cm$^{-1}$. The authors claimed that the split spectrum arises as the result of weak Joule dissipation. The spectrum of $v_{z}(t)$ within the free shear layer exhibits axial flow that is of much lower energy than the other two components, confirming that the turbulence is dominated by its 2D horizontal components.

For  $R=290.4$, the spectra exhibit different features (see figure \ref{f123}(c)). In the region of the free shear layer, the $f^{-3}$ (high frequencies) and the $f^{-5/3}$ (low frequencies) power laws are still distinguishable on the azimuthal velocity. The transition frequency, $f_{tr}t_{Ha}\simeq 46$, is higher but the corresponding non-dimensional wavenumber almost remains the same, $k_{tr}\simeq 1.09$.
Within the side wall layer, by contrast, the power density spectra of azimuthal velocity (orange line) and axial velocity (blue line) in the inertial region, exhibit a scaling of the form $f^{-5/3}$.

The apparent constance of $k_{tr}$ prompts us to compare the transition frequency and the forcing scale for cases of $R\geq 121$.  As illustrated in figure \ref{f123}(d), the scaling law for the transition frequency with $R$ follows $f_{tr}t_{Ha}\simeq 0.16R$.
Since 
in all cases, $k\simeq1$, the separation between the two slopes may simply result from the forcing geometry. In this case, the separation between $k^{-5/3}$ and $k^{-3}$  may reflect the usual 2D split between an inverse energy cascade and a direct enstrophy cascade, as already found in MHD flows by \citet{Sommeria1986}. In this case, $k_{tr}$ may be interpreted as the forcing scale, at which the mean flow transfers energy to the mean flow \citep{alexakis2018_pr}.

\subsection{Secondary flow}\label{sec:3.3}

 \citet{Potherat2000} showed that the recirculations induced by Ekman pumping  significantly influence the flow. They can be identified by looking at the variations of the velocity profile in figure \ref{regimePic}(e). Although the PSM model already provides a good understanding on the Ekman pumping effect, the detailed flow information across the fluid layer and the structures on the plane parallel to the magnetic fields remain unexplored. This is one of the motivations for carrying out 3D DNS.

  \begin{figure}
  \centering
  \includegraphics[width=0.48\textwidth]{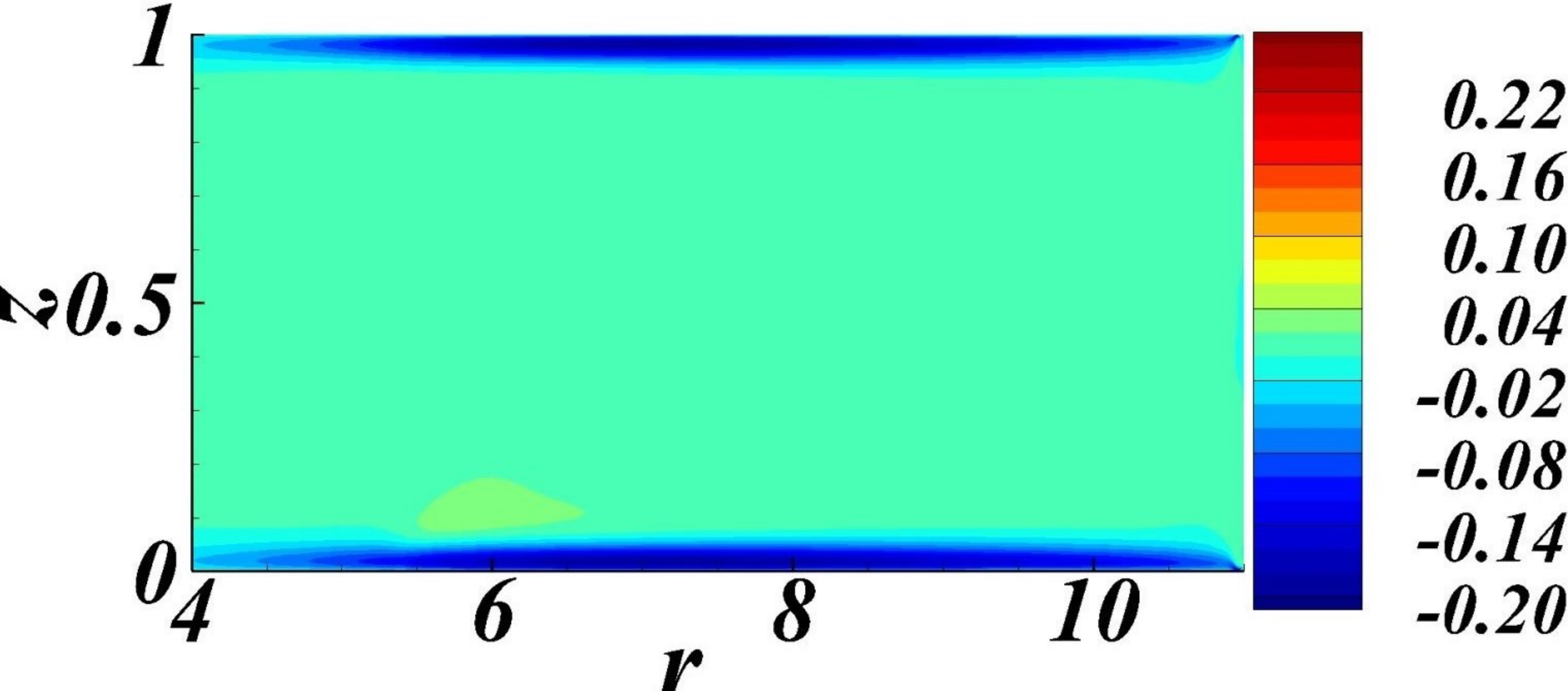}
  \includegraphics[width=0.48\textwidth]{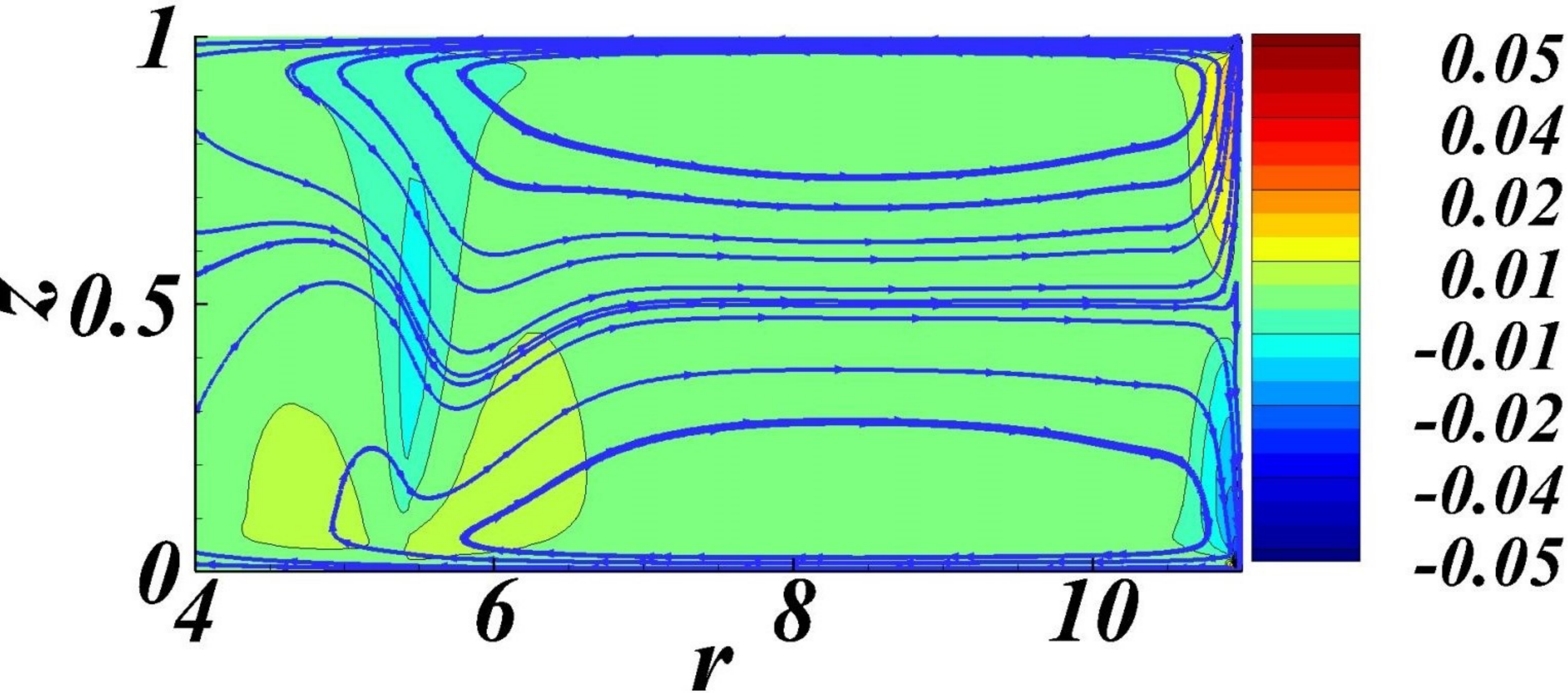}
  \caption{\label{meanEkman}The distribution of  mean radial velocity (\textit{a}), and mean axial velocity, mean streamlines (\textit{b}) on plane $\theta=0$ at $Ha=80, Re=15972$. }
\end{figure}

The streamlines of the average fluid velocity shown in figure \ref{meanEkman}(b) reveal that the Ekman pumping induces a centripetal flow ($i.e.$ away from the side wall) in the Hartmann layers and a centrifugal flow ($i.e.$ towards the side wall) in the core. In addition, the streamlines are almost symmetric with respect to the centre plane  $z=0.5$ and the recirculations consist of two symmetrical cells. Due to mass conservation, the mass fluxes in the axial direction match
the radial ones. Therefore, two strong vertical jets emerge within the wall side layer because of the decreased boundary thickness.  They flow upwards to the top Hartmann layer or downwards to the bottom Hartmann layer, where they pass through a section that is further reduced, since Hartmann layers are thinner than wall side layers. This further enhances the intensity of radial jets driven in the Hartmann layer, as shown in figure \ref{meanEkman}(a). They then turn to towards the core in the much wider region near the electrodes, into much weaker axial flows than those within the side layer, as shown in figure \ref{meanEkman}(b). This mechanism explains that recirculations are stronger within the free shear layer and the side wall layer. Ekman pumping incurs a net centrifugal transport of angular momentum as the velocity is smaller in the Hartmann layer. This has two consequences: a squeeze of the side wall layer and an increased dissipation at the side wall layer when these recirculations are important.

  \begin{figure}
\centering
\includegraphics[width=0.48\textwidth]{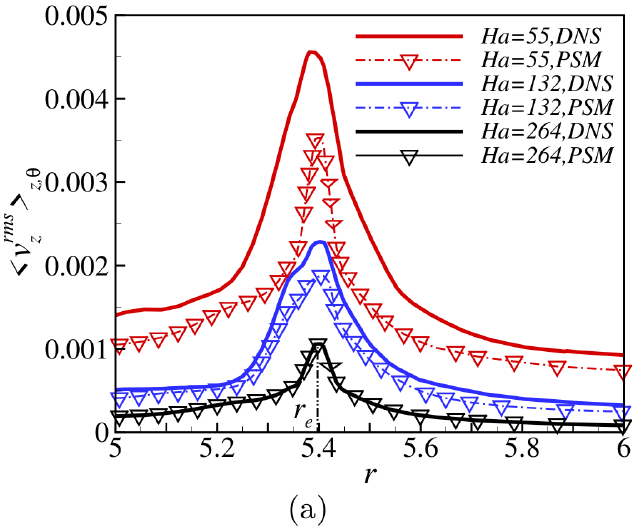}
\includegraphics[width=0.48\textwidth]{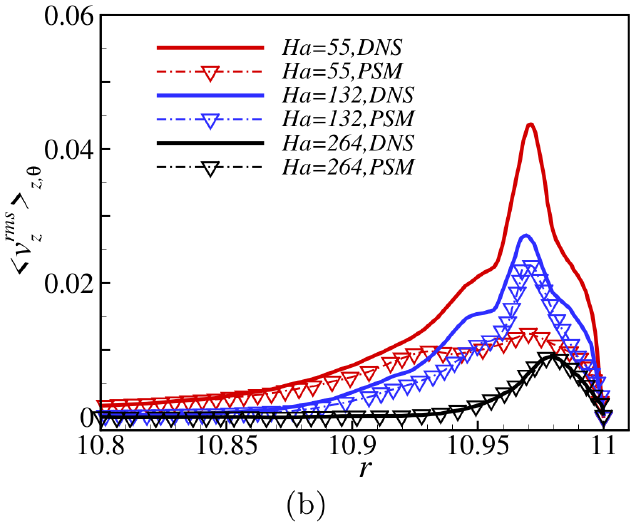}
\includegraphics[width=0.48\textwidth]{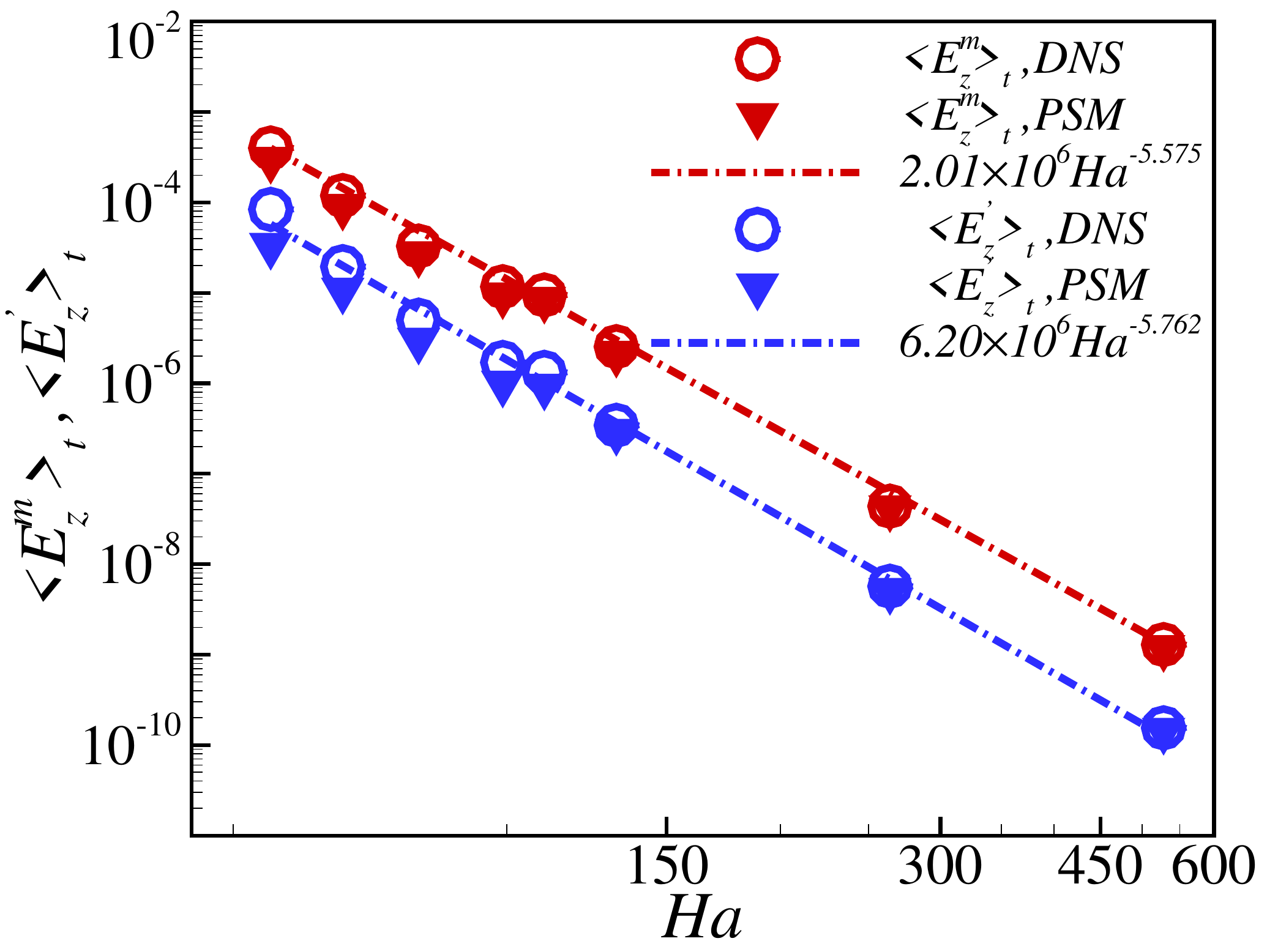}
\includegraphics[width=0.48\textwidth]{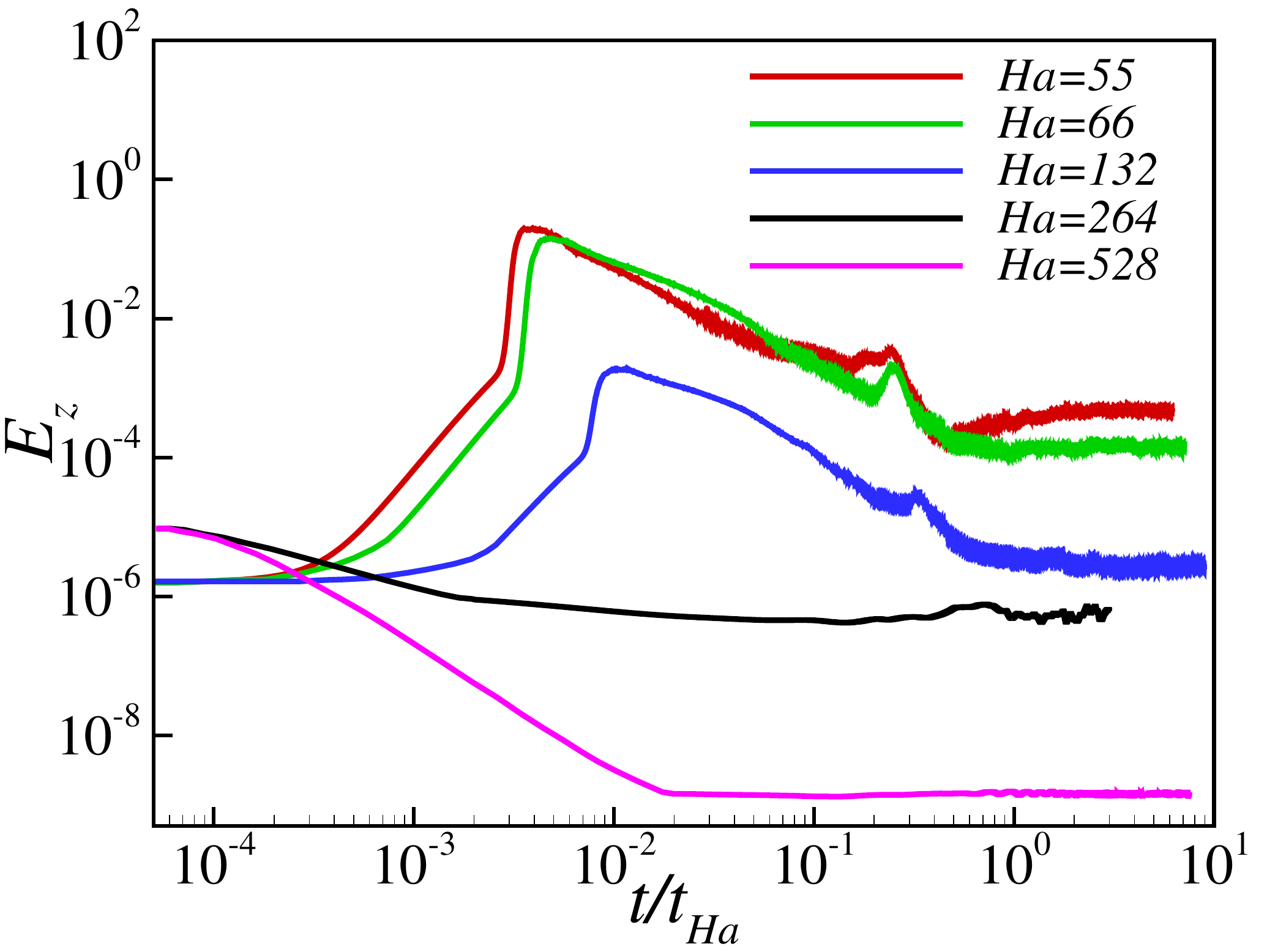}
\caption{\label{UrzandEtor}  Comparison of the $rms$ (root mean square) of the fluctuations of the vertical velocity ($u^{\rm rms}_{z}(r,\theta,z)$) along the radial direction near the side wall layer with $Re=15972$ (\textit{a}) near the current injection position $r=r_{e}$ (\textit{b}) near the side wall. Values were averaged along $\theta$ and  $z$. (\textit{c}) Average part $E^{m}_{z}=\int \langle u_{z}\rangle_t ^{2} dv$ and fluctuating part $E^{'}_{z}=\int (u^{'}_{z})^{2} dv$ of the energy in the $z-$ velocity component
(\textit{d}) Evolution of $E_{z}$, where $E_{z}=E^{m}_{z}+E^{'}_{z}$. $Re$ is fixed at 15972 for all 4 figures.}
\end{figure}

\citet{Potherat2000} have derived the analytical expression for the vertical velocity  at the interface between the Hartmann layer and the core,
\begin{equation}
  u_{z}^-=-\frac{5}{6}\frac{\lambda}{HaN}\nabla_{\bot}\cdot [(\textit{\textbf{u}}_{\bot}^{-}\cdot\nabla_{\bot})\textit{\textbf{u}}_{\bot}^{-}],
  \label{Eq:psmUz}
\end{equation}
Here, $\textit{\textbf{u}}^{-}$ denotes the velocity at the edge of the Hartmann layer and the subscript  $_{\bot}$ denotes the vector projection in the direction perpendicular to the magnetic field. Under the assumptions of PSM, any vertical component of velocity is associated to recirculations, whether local or global. Therefore, to
assess the limits of the PSM approach, we compare the energies associated to the vertical velocity component obtained with DNS at $Re=15972$ to the energy obtained from (\ref{Eq:psmUz}), distinguishing energies $E_z^m$ and $E_z^\prime$ associated to the average flow and to the fluctuations. The DNS results show that $<E_{z}^m>_{t}\sim Ha^{-5.58}$ and $<E^{'}_{z}>_{t}\sim Ha^{-5.76}$ (see figure \ref{UrzandEtor}(c), bearing in mind that $Re$ is kept constant in these scalings).
Values of $E_z$ obtained by integrating (\ref{Eq:psmUz}) from the results of PSM simulations follow the same scaling even for values of $N$ as low as $N\simeq 0.19$ ($Ha=55$) where the model is expected to break down. This indicates that PSM predicts the global recirculations associated to the mean flow very accurately.
The energy associated to the fluctuations predicted by PSM, by contrast, only matches DNS precisely for $N\gtrsim1$ ($Ha\gtrsim126.4$). Below that point, PSM underestimates $E_z^\prime$ considerably. The origin of the discrepancy can be found in the radial profiles of vertical velocity fluctuations (figure \ref{UrzandEtor}(a), figure \ref{UrzandEtor}(b)): the discrepancy between the profiles obtained with PSM and DNS  is exclusively concentrated in the free shear layer and the wall side layer. More specifically, while this discrepancy grows continuously as $Ha$ decreases  but remains moderate in the shear layer, the two profiles brutally depart from one another in the wall side layer for $Ha=55$, when small scale turbulent fluctuations appear.  Since the profiles of $\langle u_z^{\prime 2}\rangle ^{1/2}$ obtained with either PSM or DNS match everywhere else, even at for $N\lesssim0.19$, it can be concluded that PSM remains robust at predicting both global and local recirculation down to $N\lesssim1$, but breaks down when small scale turbulent fluctuations not driven by Ekman pumping appear.

Figure \ref{UrzandEtor}(d) shows the detailed evolution of $E_z$. One can see that after the injection of the electrical current density at $t/t_{Ha}=0$, part of the energy is converted to the kinetic energy along the magnetic field lines,  
which quickly results in a maximum $E_z$.  After that, $E_z$ tends to decrease until a constant time-averaged value is reached.  $E_z$ reaches a lower constant value when stronger magnetic fields are imposed for flows initialised in turbulent states.
Since the entire secondary flow transits through thin parallel layers, some residual axial flows are always observed within these boundary layers. Therefore, $E_{z}$ always stabilises at a non-zero constant value in all the numerical cases, even though it is very small at $Ha=528$ ($E_{z}<10^{-8}$). Interestingly, the walls have opposite effects on the energy in the third component, depending on whether it is driven by recirculations or turbulence: Here the residual value of $E_z$ being mostly due to Ekman pumping, it is driven by friction at the Hartmann walls. On the other hand, when turbulence freely decays in the presence of solid Hartmann walls, the energy in the third component associated to random fluctuations vanishes (in the sense that $E_z/E\rightarrow 0$ as $t\rightarrow\infty$, \citep{pk2015_jfm}). In unbounded or periodic domains, by contrast turbulence decays to a state where $E_z/E=1/2$ \citep{moffatt1967_jfm,schumann1976_jfm}.

\subsection{Boundary layers}\label{sec:3.4}
\begin{figure}
  \centering
\includegraphics[width=0.485\textwidth]{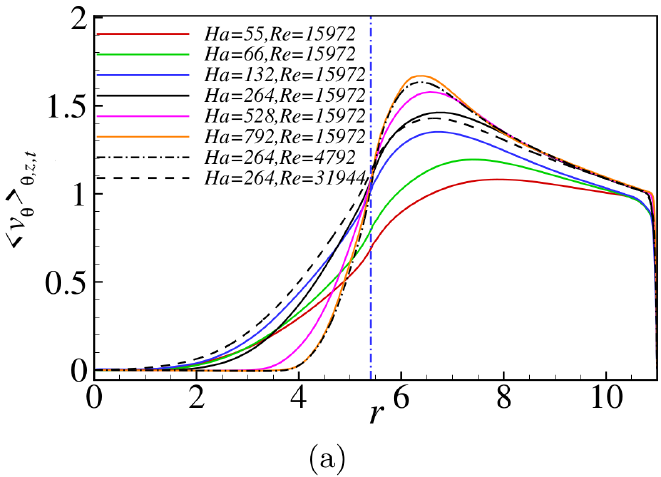}
\includegraphics[width=0.485\textwidth]{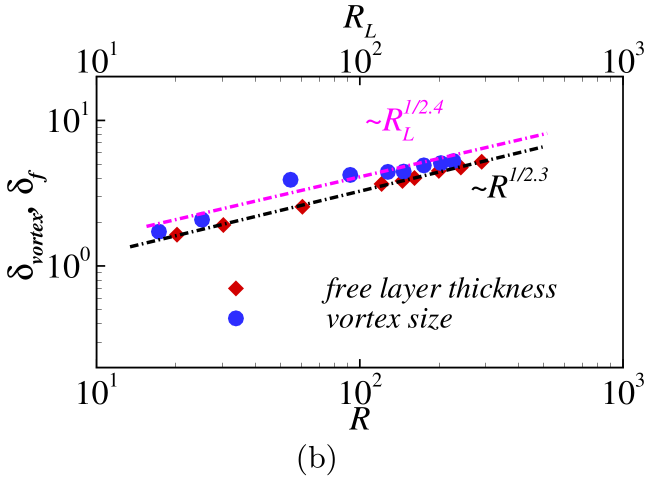}
\caption{\label{thicknesssidelayer}  (\textit{a}) Radial profiles of $\langle U_{\theta}\rangle_{\theta,z}$. (\textit{b}) Log-log plots of the free turbulent shear layer thickness $\delta_{f}$ and the large scale vortex size $\delta_{vortex}$.}
\end{figure}

One of the main purposes of the experimental study conducted by \citet{Messadek2002} was to investigate the thickness of the free shear layer when turbulence is well established. Figure \ref{thicknesssidelayer}(a) shows the radial distribution of the mean azimuthal velocity, from which the thickness of the free shear layer could be estimated. As mentioned in section \ref{sec:3.2}, the free shear layer develops quickly to an unstable state as soon as the current is injected. Therefore, a fraction
of the momentum is conveyed from the annulus to the inner region and the boundary layer thickness increases visibly.
The resulting entrainment of the fluid in the inner domain is characterized by a lower
maximum value of $U_{\theta}$ compared to that predicted by the laminar theory. For the annulus region, the (non-dimensional) value of $\langle u_{\theta}\rangle_{\theta,z,t}$ increases with $Ha$ and decreases with $Re$ due to different Ekman pumping effects. To be more specific, at moderate $Ha$, either increasing $Re$ or decreasing $Ha$  enhances the recirculations, and thus the energy dissipation is expected to grow. Conversely, for the inner region, the radial transfer of the momentum associated with recirculations set the fluid in rotation. Thus, the value of $\langle u_{\theta}\rangle_{\theta,z,t}$ in that region increases when recirculations become stronger. In other words, the thickness of the free shear layer increases  when the Ekman pumping effect becomes stronger, which is opposite to their effect on the laminar side layer ($Ha^{-1/2}$). 
%

We now use the mean azimuthal velocity profiles within a wide parameter spaces of $\{Re,Ha\}$  to determine the thickness of the shear layer $\delta_{f}$, which is defined as,
\begin{equation}
 \delta_{f}=\frac{\Delta U_{\theta}}{(dU/dr)_{max}}.
\end{equation}
Here, $\Delta U_{\theta}=U_{\theta max}-U_{\theta min}$, $U_{\theta min}=0$ and $U_{\theta max}$ is the mean velocity value at the
intersection of the maximum slope and the mean velocity profile predicted by the laminar theory (\ref{Eq:Uteta1}). As suggested by \citet{Messadek2002}, the layer thickness $\delta_{f}$ depends on both $Re$ and $Ha$ according to the relation
\begin{equation}
 \delta_{f}=C_{f}(R)^{1/n}.
\end{equation}
As shown in figure \ref{thicknesssidelayer}(b), the best fit from our data yields $C_{f}\simeq0.44$ and $n\simeq2.3$ respectively, which is consistent with the values obtained  by \citet{Messadek2002}. The maximum relative error between the fitting curve and the calculation data is lower than $5.1\%$. Moreover, for $(Ha,Re)=(792,15972)$, the layer thickness from numerical simulation, $\delta_{f}\approx1.64$, agrees well with the experiment one, $\delta_{f}\approx1.61$.
This scaling is very different from the theoretical laminar one ($\delta_{f} = Ha^{-1/2}$, independent of $Re$) and reflects the role of two-dimensional inertia, measured by $R$, in determining the thickness of the free shear layer.
In addition, the size of large scales  estimated from the  velocity fluctuations (see \citet{Potherat2005}) is also shown. Here, $R_{L}$ represents the Reynolds number based on the velocity and the size of the large vortices and he size of the  large vortices is estimated from the profiles of \emph{rms} of azimuthal velocity fluctuations. As shown in figure \ref{thicknesssidelayer}(b), their size follows a very similar scaling to the size of the boundary layer $\delta_v\simeq 0.57 R_L^{2.4}$ with a maximum relative discrepancy to that law lower than $5.6\%$.

\begin{figure}
\centering
\includegraphics[width=0.6\textwidth]{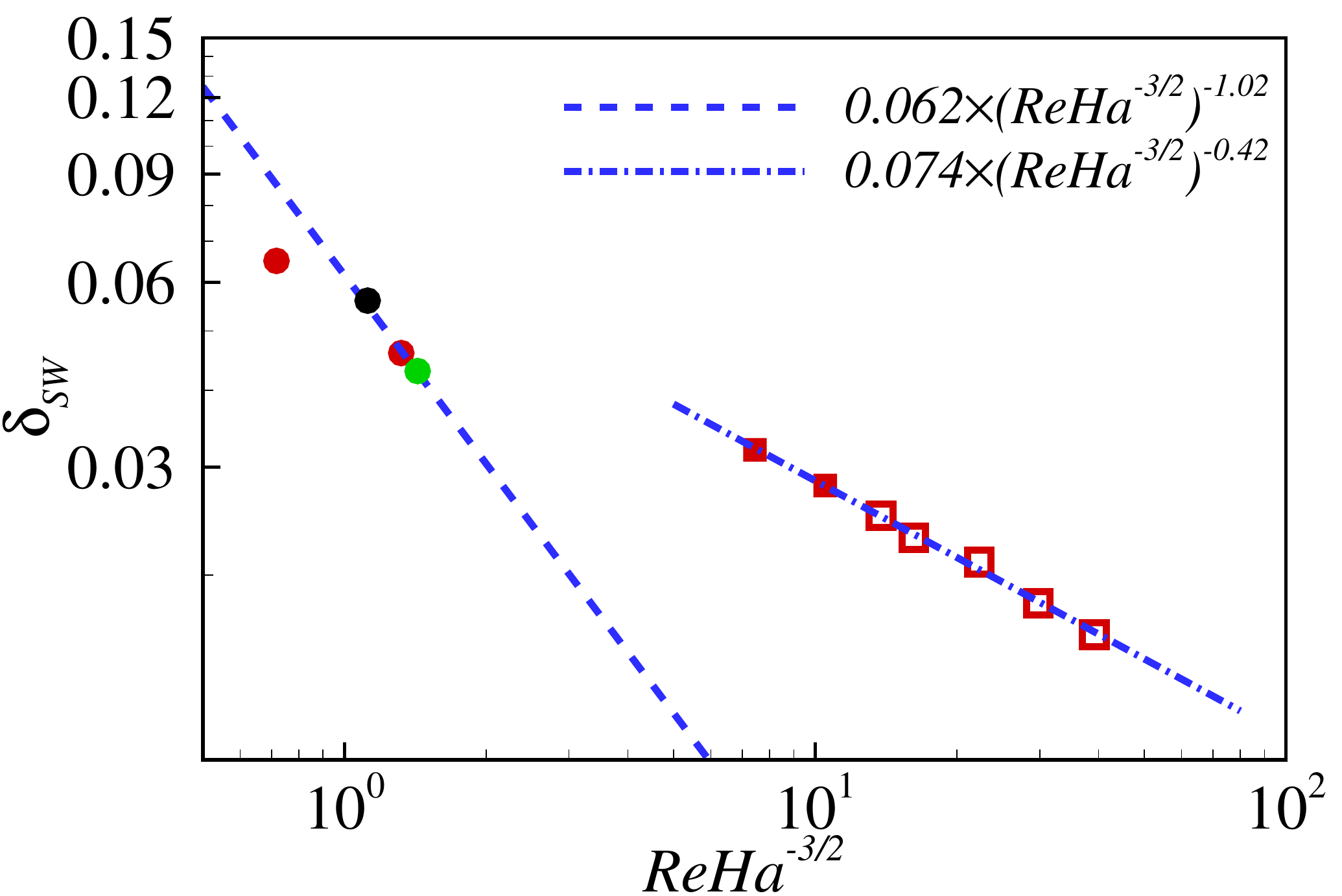}
\caption{Variations of the wall side layer thickness $\delta_{\rm SW}$ against parameter $C=ReHa^{3/2}$ \citep{Potherat2000} for $Re=15972$ (red), $Re=31944$ (green) and $Re= 4792$ (black). Full circles: boundary layer in laminar state, full squares: attached boundary layer with 3D turbulence, hollow squares: separated boundary layer. \label{fig:wallside_thickness} }
\end{figure}

Unlike the free shear layer, the structure of the wall side layer was not experimentally accessible. Three-dimensional simulations make it possible to examine how its thickness varies against parameter $C=ReHa^{-3/2}$, which \citep{Potherat2000} identified as the governing parameter when secondary flows dominate. The thickness $\delta_{\rm SW}$ was defined as the distance from the wall to the point where the velocity magnitude reaches 90$\%$ of $U_\theta^{\rm SW}$, where $U_\theta^{\rm SW}$ is the velocity  value at the intersection of the minimum slope and the mean azimuthal velocity profile. 
The results are reported on Figure \ref{fig:wallside_thickness}. For low values of $C$, recirculations are weak and $\delta_{\rm SW}$ is expected to scale as $\delta_{\rm SW}\sim Ha^{-1/2}$. As $C$ increases, recirculations become more prominent and
$\delta_{\rm SW}$ approaches the theoretical scaling of $\delta_{\rm SW}\sim C^{-1}$. The picture changes slightly before (in the sense of increasing $C$) three-dimensional turbulence appears in the boundary layer: the thickness suddenly increases to settle on a larger scaling
characterised by $\delta_{\rm SW}\simeq 0.074(Re Ha^{-3/2})^{-0.42}$. More simulations would be needed to confirm that $C$ remains the relevant parameter in this regime (though the continued prominence of the secondary flows would suggest this may be the case) and to confirm the exponent of $Re$ in this scaling. Interestingly, boundary layer separation has little visible impact on the scaling of $\delta_{\rm SW}$, most likely because of the small ratio of the surface where it occurs to the total surface of the wall.
\subsection{Angular momentum and wall shear stress}\label{sec:3.5}
For a first estimate the the global angular momentum, we note that most of the viscous and Joule dissipation takes place within the Hartmann layer for Q2D
flows under a strong magnetic field.
Since the most intense part of the flow occurs in the outer
annulus where the driving force acts, the contribution of the inner region to the total angular momentum can be neglected to derive the simple expression (\ref{Lameqn})  from the theory of \citet{Sommeria1982} for the elementary case of a steady inertialess flow.
Note that in this case, the Hartmann layers remain laminar and inertialess. This implies that the angular
momentum varies linearly with $I$ and is independent of $B$.

 \begin{figure}
  \centering
  \includegraphics[width=0.6\textwidth]{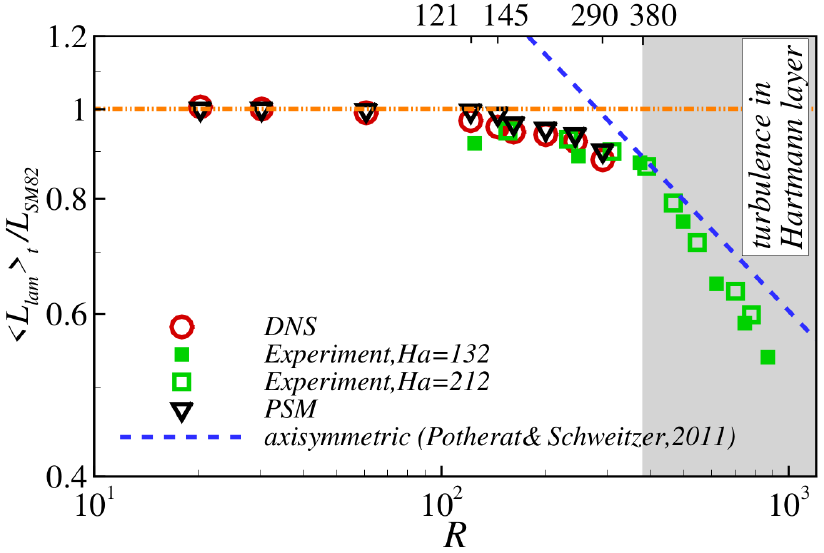}
  \caption{\label{Lam82}Variations of the global average angular momentum $\langle {L}_{lam}\rangle_t/L_{SM82}$ versus $R$. Region with $R\geq380$  \citep{Moresco2004} is defined as regime where turbulence formed within the Hartmann layer.  Region with $145.2\leq R\leq 290.4$ is defined as flows with turbulence emerging within the side layer. Region with $R < 121$ is corresponding to flows without separation of side layer.}
\end{figure}

The values of $L_{lam}$ obtained from the present numerical results (circle open symbols) are plotted in figure \ref{Lam82}, along
with the values of the angular momentum measured in MATUR \citep{Messadek2002} and predicted with PSM model (triangular open symbols). All the data reported in this figure is normalised by the value $L_{SM82}$ predicted with the theoretical expression (\ref{Lameqn}), the dashed line corresponds to the theoretical prediction derived by \citet{Potherat2011} for turbulent Hartmann layers and the square symbols denote the experimental data. As shown in figure \ref{Lam82}, the experimental values, the results of PSM model and our numerical values  collapse well into a single curve, which can be divided into three different zones demarcated by changes in slope around $R\simeq 121$ and $R\simeq 380$. When $R < 121$, the numerical values (DNS and PSM) are almost equal to unity, which matches the  SM82 linear approximation closely. For $R \geq 380$, the experimental  values fall to significantly lower values than the linear prediction as would be expected when turbulence arises within the Hartmann layers. However, they match  well with the  values predicted with the simplified axisymmetric model derived by \citet{Potherat2011}, which supposes that the Hartmann layers are turbulent. When  $121 \leq R< 380$, a relatively large discrepancy can be observed between the numerical values (or experimental values) and the theoretical value of SM82 or theoretical value derived by \citet{Potherat2011}.

An explanation for this intermediate range was provided by \citet{Potherat2000}, who modified the equation for $L_{lam}$ from the SM82 theory to account for the dissipation in the side layers,
\begin{equation}{\label{Lamtotaleqn}}
   \frac{d L_{lam}}{dt}=F-S_{\nu}-\frac{2L_{lam}}{t_{Ha}},
\end{equation}
where $F$ denotes the global electric forcing  and $S_{\nu}$ ($\sim \overline{\tau^{Sh}}$) denotes the viscous dissipation at the side wall layer. At a small forcing, the corresponding viscous effect on the angular momentum is negligible in comparison with the Hartmann friction. Thus, the DNS and PSM results are consistent with the values predicted with the theoretical expression (\ref{Lameqn}) when $R<121$. By contrast, significant differences emerge between the numerical values  and the SM82 theory when $121 \leq R<380$, since in this range, the viscous dissipation at the side wall layer cannot be ignored any more. Because the side wall layer is squeezed by the strong Ekman pumping recirculations, the side wall layer becomes thin, which results in significant increase of velocity gradient and the global dissipation. The present numerical results  confirm this conclusion well when $R\in[145.2, 290.4]$, for which these recirculations are significant.

 \begin{figure}
  \centering
  \includegraphics[width=0.6\textwidth]{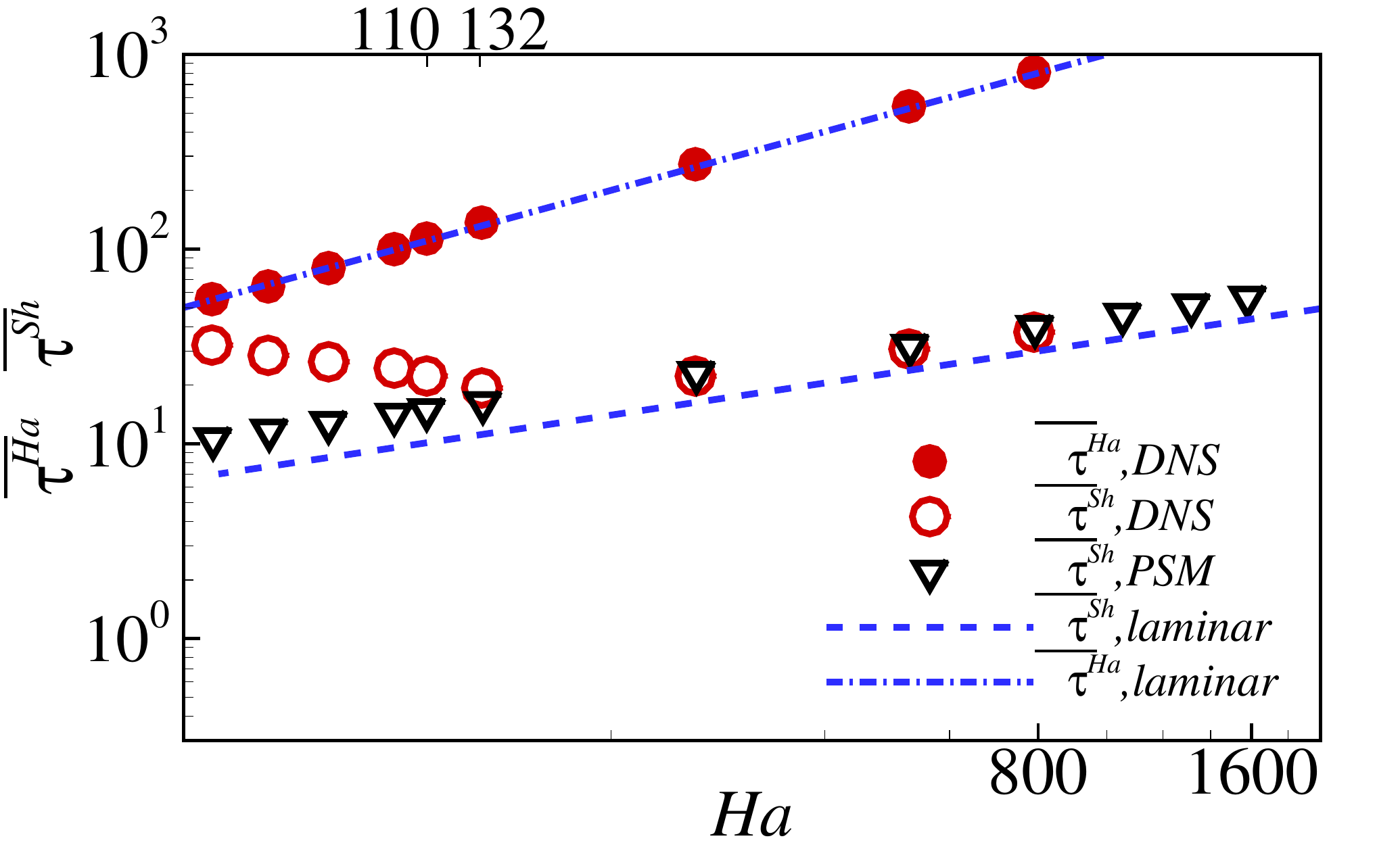}
  \caption{\label{tauside_R} Space and time-averaged wall friction over the entire range of $Ha$ investigated and for $(Re=15972)$: circle full symbols denote wall stress  at Hartmann wall and circle opened symbols denote wall stress on the side wall For comparison, theories for laminar flows (dashed and dashed-dot line) and results of the PSM model (gradient opened symbols) are shown too.}
\end{figure}

The DNS enable us to go a step further and examine in detail how the wall shear stress is affected three-dimensional effects. The variation of the wall stresses ($\overline{\tau^{Ha}}$, $\overline{\tau^{Sh}}$) with $Ha$ are shown in figure \ref{tauside_R}.
 The values of wall stress are different on the Hartmann walls and side walls. As shown in  figure \ref{tauside_R}, all the shear stresses on the Hartmann walls collapse well into a single curve (laminar solution, $\overline{\tau^{Ha}}\sim 2Ha/Re$), even when $Ha=55$, which indicates that the Hartmann layers remain laminar for all the cases considered  in this work. However, for the side wall shear stresses, the values gradually depart from the laminar solution as $Ha$ decreases ($Ha\leq132$).  To understand the role played by recirculation in this phenomenon,  all the cases are rerun  with  PSM model, as well as additional cases with higher $Ha$ (1056, or 1320, or 1584). When $Ha\geq 264$, one can see that the results of DNS match well with that of 2D simulations, but they are noticeably higher than the straight duct laminar wall stresses  because of the recirculations induced by Ekman pumping. For cases with $Ha\leq 132$, the recirculations become more and more significant as $Ha$ decreases, which results in an increased squeezing of the side wall layer. Therefore, the values of $\overline{\tau^{Sh}}$ obtained from PSM
simulations are higher than those predicted by the scaling law for a straight laminar Shercliff layer. However, DNS values of $\overline{\tau^{Sh}}$ are still significantly higher than the ones from the PSM  for $Ha\leq132$. This coincides with the observation that PSM cannot
capture the small-scale turbulence within the side layers in this range of parameters and further indicates that the dissipation it incurs
dominates even the enhanced dissipation due to the squeezing of the side layer by secondary flows.
In summary, the DNS confirm that the Hartmann layer remains laminar for cases with $R \leq 290.4$. When $121 \leq R\leq 290.4$, the discrepancy between the numerical  results and the theoretical results is associated with either the strong squeezing of the side wall layer or turbulence within side wall layer.

  \begin{figure}
  \centering
\includegraphics[width=0.48\textwidth]{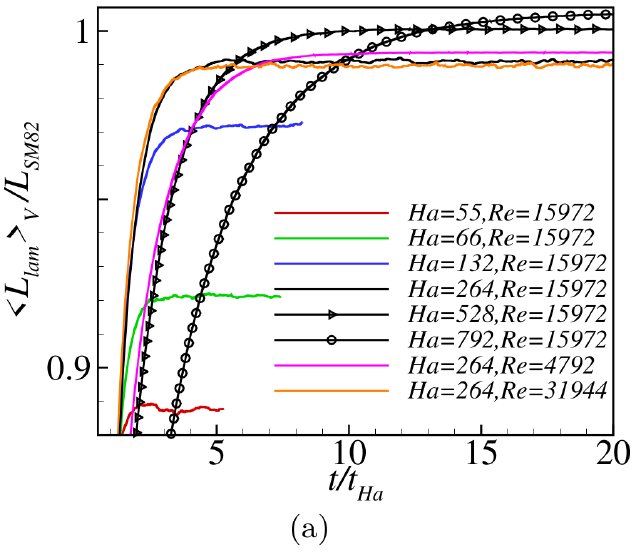}
\includegraphics[width=0.48\textwidth]{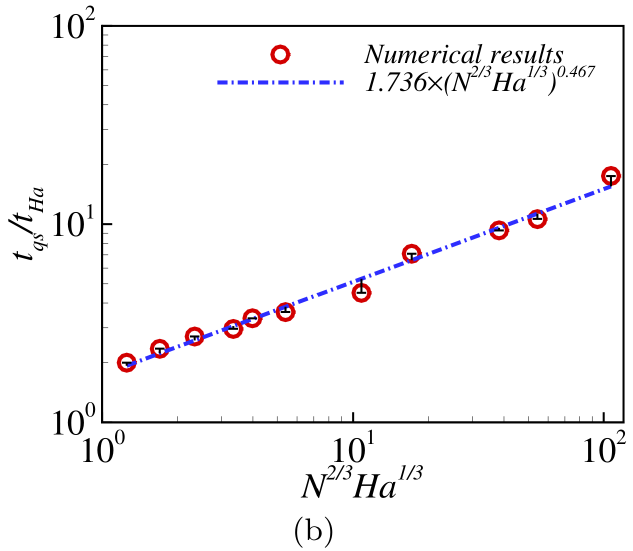}
\caption{\label{TransientTime}(\textit{a}) Evolution of the global angular momentum with time. (\textit{b}) Transient time obtained numerically after switching on the forcing on a fluid at rest versus the non-dimensional parameter $N^{2/3}Ha^{1/3}$.} 
\end{figure}
Finally, we compare the time variations of the mean angular momentum for different $Re$ and $Ha$, 
 shown in figure \ref{TransientTime}(a). For the same $Ha$, one can see that the amplitude of the oscillations of the
angular momentum in the quasi-steady-state increase with $Re$, due to the increasingly turbulent nature of the flow.
 Furthermore, the transient time ($t_{qs}$) for the system to reach the quasi-steady state from the fluid being at rest
decreases with  $Re$ (we estimate this time by measuring the slope of the
$L_{lam}(t)$ curve near equilibrium in a log-log diagram, as in \citet{Potherat2005}). This too is
associated with global dissipation, which is significantly increased by the strong Ekman pumping at higher $Re$, shortening the transient time. For a given $Re$, the variation trend with $Ha$ is opposite, reflecting the damping of recirculations by the Lorentz force.
This effect is quantified in Figure \ref{TransientTime}(b), which shows the variation of the transient time with the combined non-dimensional parameter $N^{2/3}Ha^{1/3}=Ha^{5/3}/Re^{2/3}$, with a scaling  $t_{qs}/t_{Ha}\simeq1.736(N^{2/3}Ha^{1/3})^{0.467}$, with a maximum relative error between the fitting curve and the numerical data of $9.1\%$. While $t_{qs}/t_{Ha}$ is governed by the same parameter as predicted by PSM, the scaling exponent of 0.467 stands much lower than the value of 1, indicated in \citet{Potherat2005}. Since a lower exponent is indicative of a higher level of dissipation, this discrepancy may be attributed to the dissipation incurred by 3D turbulence in the side layers that the PSM model cannot account for.

\subsection{Three dimensionality}\label{sec:3.6}

According to \citet{Potherat2014} and \citet{Potherat2015}, the dimensionality of a structure in a channel of gap $a$
is determined by the ratio $l_{z}/a$. Here $l_{z}$ is the momentum diffusion length along $\textbf{\textit{B}}$  by the Lorentz force.
This diffusion process takes place in a typical diffusion time $\tau_{2D}(l_{\bot})=\tau_{j}(l_{z}/l_{\bot})^{2}$, where $\tau_{j}=\rho/\sigma B^{2}$ is the Joule dissipation time.
Hence, $l_{z}$ can be estimated as \citep{Sommeria1982}
\begin{equation}
l_{z}(l_{\bot})=l_{\bot}\sqrt{N(l_{\bot})},
\end{equation}
where $N(l_{\bot},u(l_{\bot}))=\sigma B^{2}l_{\bot}/\rho u(l_{\bot})$, is a scale-dependent interaction parameter and $u(l_{\bot})$ is the velocity associated to a fluid structure of size of $l_{\bot}$.  $l_{z}/a\gg 1$ indicates that the Lorentz force diffuses its momentum over a distance much greater than $a$, and that the considered structure is consequently quasi-two dimensional.

\begin{figure}
  \centering
\includegraphics[width=0.32\textwidth]{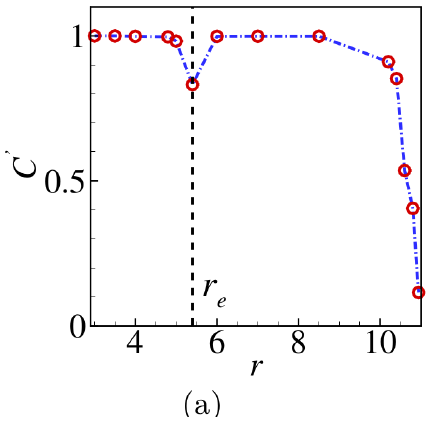}
\includegraphics[width=0.32\textwidth]{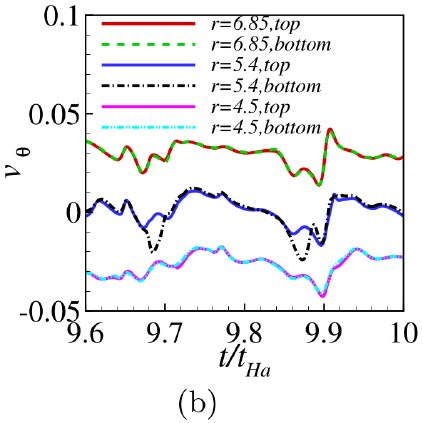}
\includegraphics[width=0.32\textwidth]{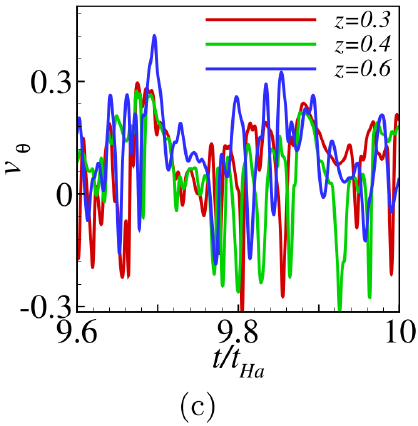}
\caption{\label{3dimensionality}The results of case with $Ha=55, Re=15972$. (\textit{a}) Radial profile of correlation $C^{'}$. (\textit{b})  The  azimuthal velocity fluctuation signals of six different points in the top and bottom Hartmann layer variation with time. For easy identification, values of $u_{\theta}$ at $r=6.85$ ($r=4.5$) is displayed  as  $u_{\theta}+0.03$ ($u_{\theta}-0.03$). (\textit{c}) The azimuthal velocity fluctuations  signals of three different points in the side wall layer variation with time, along $r=10.8$.}  
\end{figure}

However, the separation of the side wall may produce complex 3D structures in the case of $Ha=55, Re=15972$. Therefore, the correlation between $V_{T}(r,\theta,t)$ and $V_{B}(r,\theta,t)$ measured at locations within either Hartmann layers exactly aligned with the magnetic field lines
(i.e. respectively at $z=z_{0}$ (with $z_{0}=0.008 < 1/Ha$) and $z=1-z_{0}$, but at the same coordinates $(r, \theta)$) is used to assess the three dimensionality of the flow \citep{KLEIN2010}. Here, $V_{T}(r,\theta,t)$ and  $V_{B}(r,\theta,t)$ represent the azimuthal velocity fluctuations. The correlation function is defined as
\begin{equation}
   c^{'}(r,\theta)=\frac{\sum\limits_{t=0}^{Ti} V_{T}(r,\theta,t) V_{B}(r,\theta,t)}{\sum\limits_{t=0}^{Ti} V_{B}^{2}(r,\theta,t)}.
\end{equation}
where $Ti$ is the duration of the recorded signals. Considering the symmetry of the problem, we shall analyse the radial dependence of this 
correlation through $C^{'}(r)=\langle c'(r,\theta)\rangle_\theta$.
For $Ha=55, Re=15972$, the correlation profile of $C'(r)$ in figure \ref{3dimensionality}(a) indicates  that the flow is very close to quasi-2D in most regions, where correlation $C^{'}$ is almost unity. This is also demonstrated on figure \ref{3dimensionality}(b), which shows that the instantaneous azimuthal velocity signals away from the shear layers near the top and bottom Hartmann layers are near identical for the same position ($r,\theta$). At $r=r_{e}$, by contrast, the signals slightly differ,
so the correlation $C^{'}$ is slightly lower than unity (0.832). At this location, the signals are mostly identical except for a slightly
higher amplitude of the bottom signal and short burst where the signals are weakly correlated. The difference in amplitude reflects \emph{weak} three-dimensionality, as defined by \cite{KLEIN2010}, in the sense that flow near the top and bottom are identical in topology but differ
in intensity. The short bursts, by contrast, indicate \emph{strong} three-dimensionality where topologies are no more identical. The bursts
correspond to the passage of coherent structures. Hence, the overall picture is that while the free shear layer itself and the larger
structures responsible for the lower frequency oscillations are only weakly three-dimensional, the smaller coherent structures that
navigate along it can exhibit strong three-dimensionality.

Figure \ref{3dimensionality}(c) confirms that strongly three-dimensionality emerge within the side wall layer, where the signals recorded at all three monitored depths differ noticeably. This is consistent with the distribution of azimuthal velocity within the side layer shown in figure \ref{regimePic}(f), where the flows on different transverse plane are not topologically equivalent any more, because of the presence of small scale three-dimensional turbulence there.

 \begin{figure}
  \centering
\includegraphics[width=0.31\textwidth]{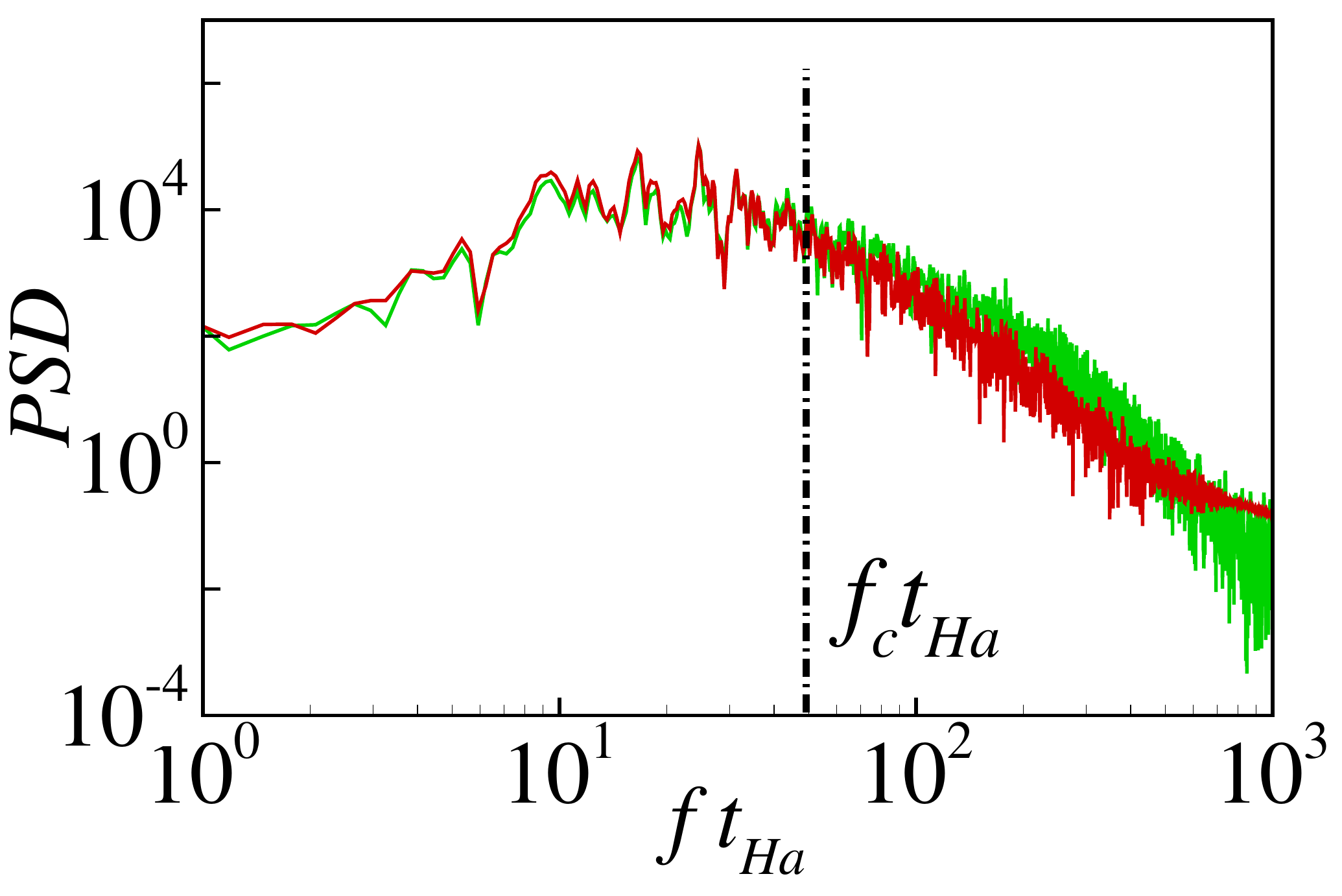}
\includegraphics[width=0.31\textwidth]{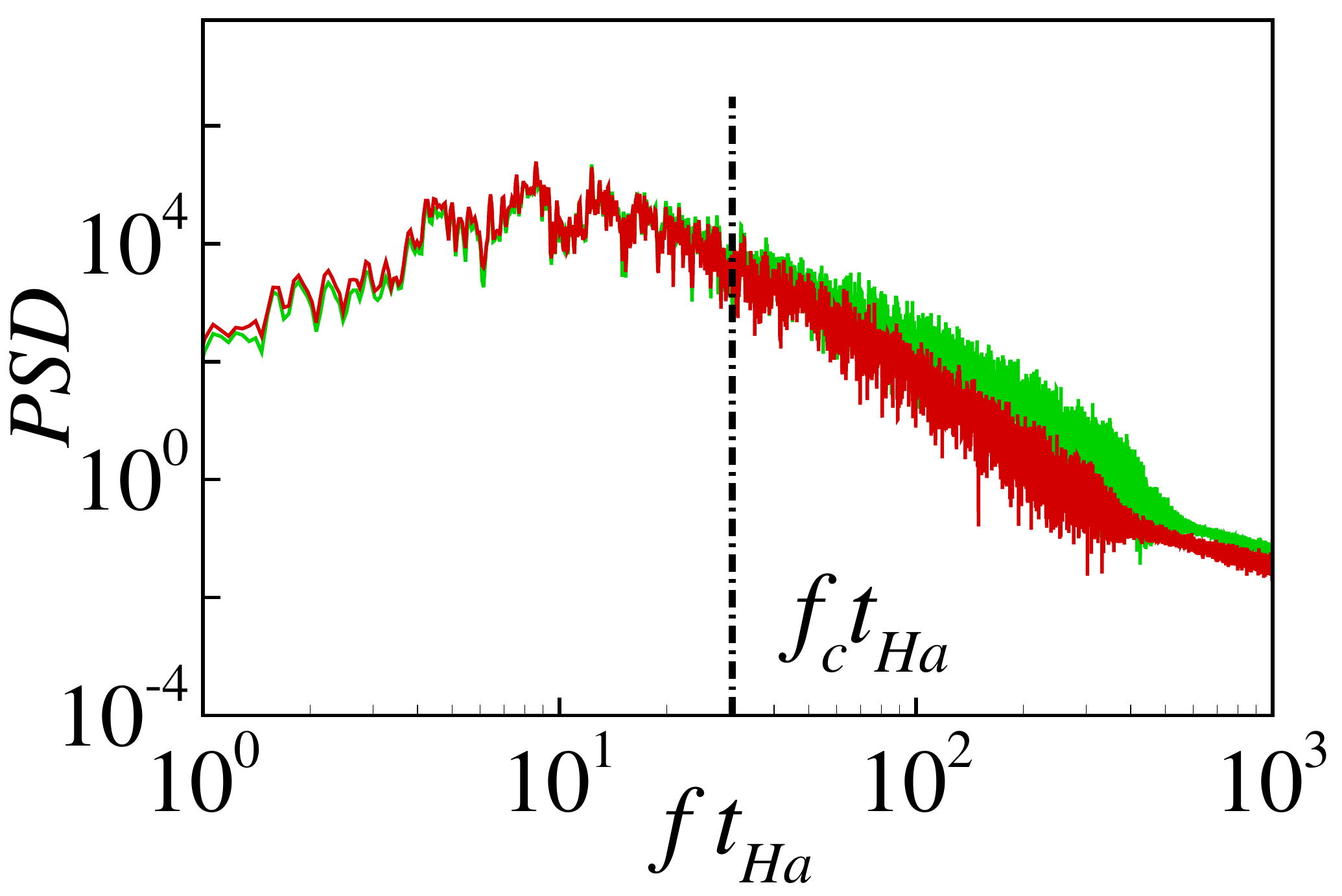}
\includegraphics[width=0.31\textwidth]{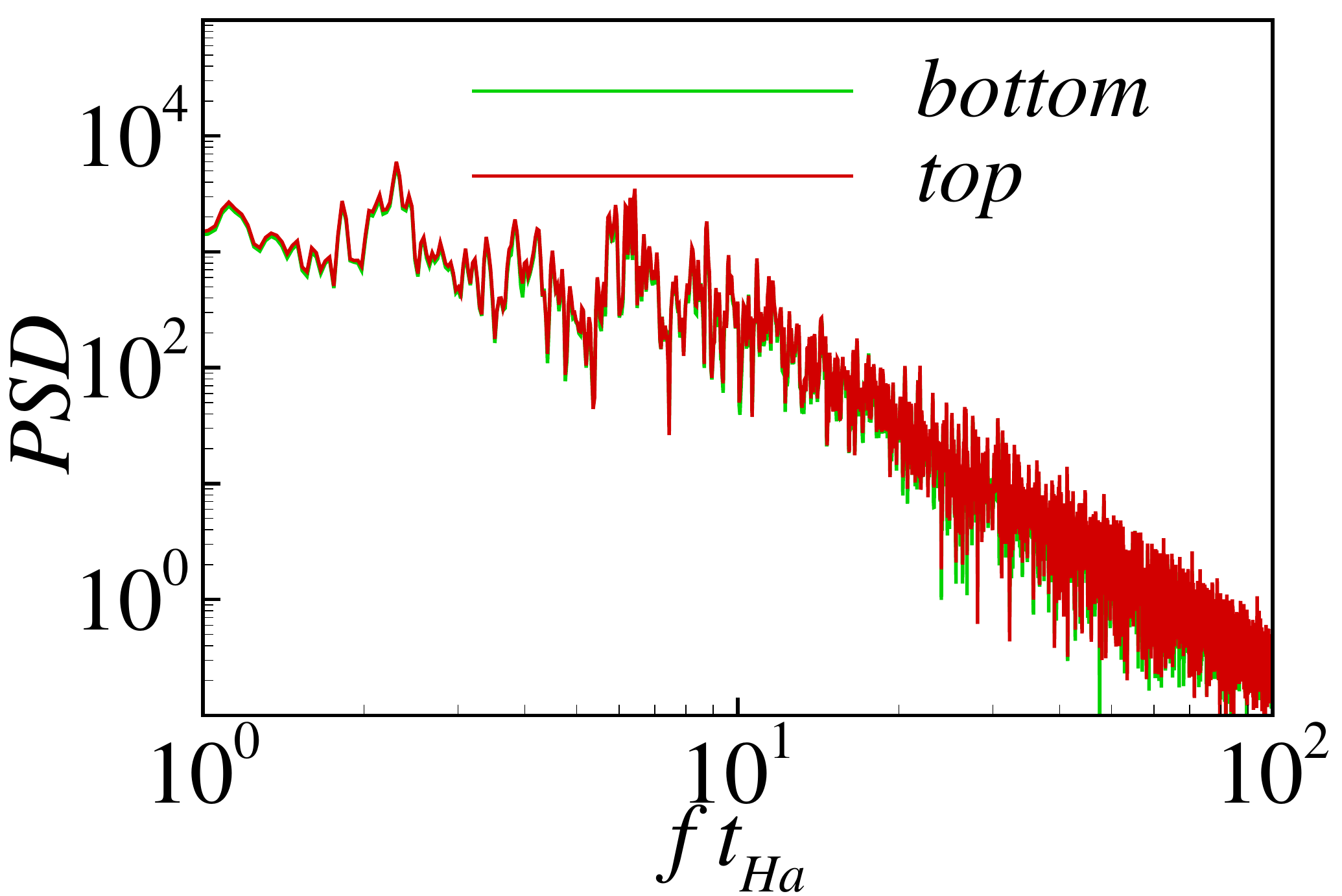}
  \caption{\label{fc123}Power density spectra calculated from  instantaneous azimuthal velocity signals acquired at locations near the free shear layer inside the top (red line) and bottom (green line) Hartmann layers for different magnetic interaction parameters. (\textit{a})  $Ha=66, Re=15972$. (\textit{b})   $Ha=264, Re=31944$. (\textit{c})   $Ha=264, Re=15972$.}
\end{figure}

To further characterise the scale-dependence of three-dimensionality near the electrodes, we analyse the spectra obtained from the velocity signals near the top and bottom Hartmann layers, as shown in figure \ref{fc123}. For $Ha=66, Re=15972$ and  $Ha=264, Re=31944$, pairs of energy spectra obtained near top and bottom Hartmann walls reveal that the higher frequencies carry significantly less energy in the vicinity of the top wall than that near the bottom wall, which is a clear evidence of three-dimensionality. By contrast, lower frequencies almost carry the same amount of energy.
According to the theory of \citet{Sommeria1982} and the experiments of \citet{Baker2018}, a cutoff scale $k_{c}$ (corresponding to $f_{c}$ here) separating the quasi-two-dimensional structures from
the three-dimensional structures can be identified. Here, the velocity signals, averaged over the azimuthal direction, were taken within the free shear layer, so as to minimise the influence of the side walls, \emph{i.e.} $r=r_{e}$. For the details about the calculation of $f_{c}$, the readers are referred to \citet{Potherat2014}. When $N_{t}$  decreases, $f_{c}$ decreases as three dimensionality contaminates larger and larger scales,
as shown in figure \ref{fc123}(a,b). For $Ha=264$ and $Re=15972$, fluctuations within the free shear layer are quasi-two dimensional over the entire spectral range (see figure \ref{fc123}(c)).
 \begin{figure}
  \centering
\includegraphics[width=0.48\textwidth]{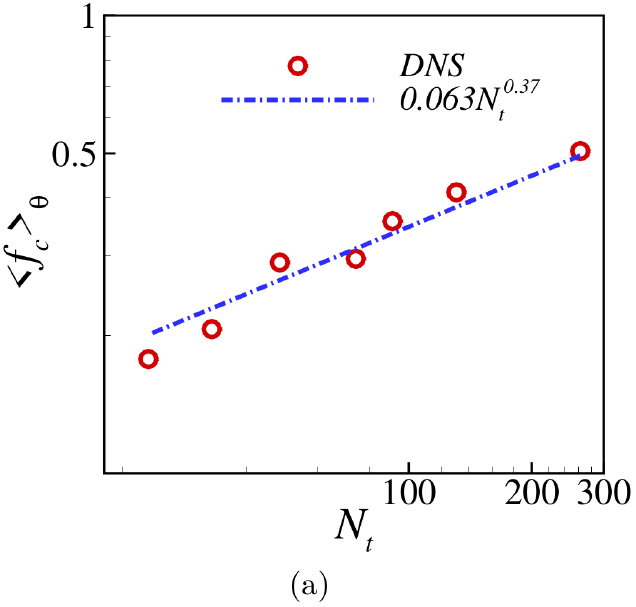}
\includegraphics[width=0.48\textwidth]{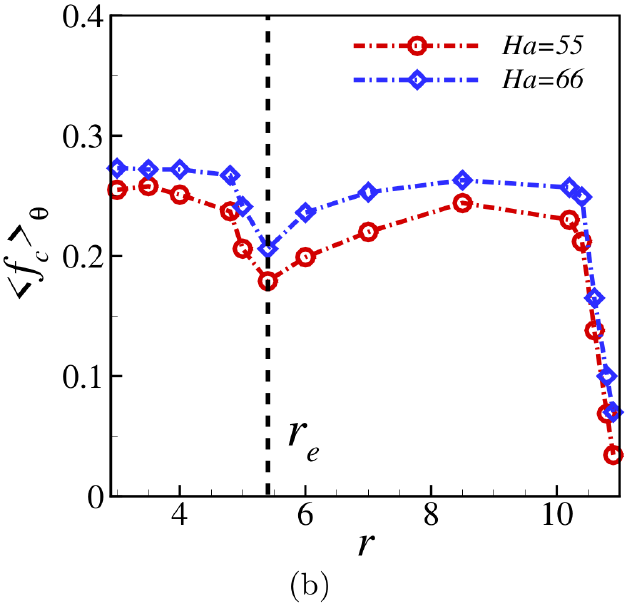}
\caption{\label{fcNt}(\textit{a})  The azimuthal averages of $f_{c}$, separating Q2D large structures from the  small 3D ones, normalised by $U_{0}/a$. (\textit{b}) Radial profiles of azimuthal averages of $f_{c}$, when $Ha=55, Re=15972$ and $Ha=66, Re=15972$.}
\end{figure}
The variations of the azimuthally averaged cut-off frequency $\langle f_c \rangle_\theta (N_{t})$ are represented in figure \ref{fcNt}(a), and reveal that the variations of $\langle f_c \rangle_\theta$ across all investigated cases collapse onto a single curve, and therefore that $\langle f_c \rangle_\theta$ is determined by $N_{t}$ with a scaling law
\begin{equation}{\label{fc}}
\langle f_c \rangle_\theta \simeq0.063N^{0.37}_{t}.
\end{equation}
This general law gives a clear estimate for the minimum frequency of vortices that are affected by 3D inertial effects. Additionally, we obtain the minimum transverse wavenumber of 3D vortices, $\langle k_c \rangle_\theta \simeq 0.396 N^{0.37}_{t}$ by applying Taylor's hypothesis, taking advantage of the strong azimuthal flow component. This law is the first numerical confirmation of the original theoretical law given by \citet{Sommeria1982}, $k_{c}\sim N^{1/3}_{t}$, following \citet{Baker2018}'s recent experimental confirmation. It is also interesting to note that while this law applies to homogeneous and sheared turbulence alike, the corresponding scaling law for the cutoff frequency (\ref{fc}) differs from the power law with exponent $\sim 2/3$ found in turbulence with weak average flow \citep{KLEIN2010}.

Given the spatially inhomogeneous nature of the flow in the radial direction, the question arises as to where three-dimensionality is preferentially found. An answer is provided by the
spatial distribution of the azimuthally averaged variations of the cutoff frequency $\langle f_{c} \rangle_\theta$ along $r$,
shown in figure \ref{fcNt}(b).
$\langle f_c \rangle_\theta$ clearly drops at the locus of the free shear layer and the side wall layer. On the other hand, it  remains much higher outside of these regions. Hence, structures are quasi-two-dimensional over a greater range of scales, down to smaller ones outside the shear layers and three-dimensional turbulence appears concentrated in the side layer and to a slightly lower extent, to the free shear layer. This suggests that three-dimensionality arises out of direct energy transfer from the mean shear flow to scales at the scale of the shear layers ($\delta$) or lower. This is supported by the argument that the high level of energy imparted to them by the shear
makes the turnover time at this scale ($\sim \delta/\langle U_\theta(r)\rangle$, with $r=r_e$ or $r=R$ for the free shear and side layers respectively) significantly smaller than the two-dimensionalisation time ($\sim (\sigma B^2/\rho)(a/\delta)^2$). Indeed, the ratio of former to the latter expresses as $N(\delta/a)^2$, which is significantly smaller than unity in both cases.

Conversely, away from the shear layers, the mean flow does not inject energy into the small scales. Since
the recent experiments on MHD turbulence without a strong mean flow of \cite{Baker2018} suggest that even in the presence of moderate three-dimensionality, the energy cascades upscale, the flow is dominated by larger vortices for which two-dimensionalisation is more efficient. Unlike in these experiments, however, the presence of vorticity streaks in the wake of these vortices suggests that an additional transfer mechanism less favourable to large scales may be at play in the outer region of MATUR.
\subsection{Componentality}
To understand the occurrence of the third component velocity fluctuations not driven by global Ekman pumping, we show instantaneous distributions of axial velocities in the cross-section $\theta=0$ (see figure \ref{structure}(a)). One can see that the turbulent fluctuations are localized within the layer near the side wall. Moreover, this three-dimensionality contaminates the entire height of the vessel, from the bottom to the top of the Hartmann layers. As conveyed in figure \ref{structure}(b,c), the distributions of instantaneous vertical velocity (including intensity and topological structure) are different on the  plane near the top wall ($z=0.8$) and near the bottom wall ($z=0.2$). This further confirms that
the wall side layer is the region where practically all strong three-dimensionality is concentrated.
%
 \begin{figure}
   \centering
  \includegraphics[width=0.45\textwidth]{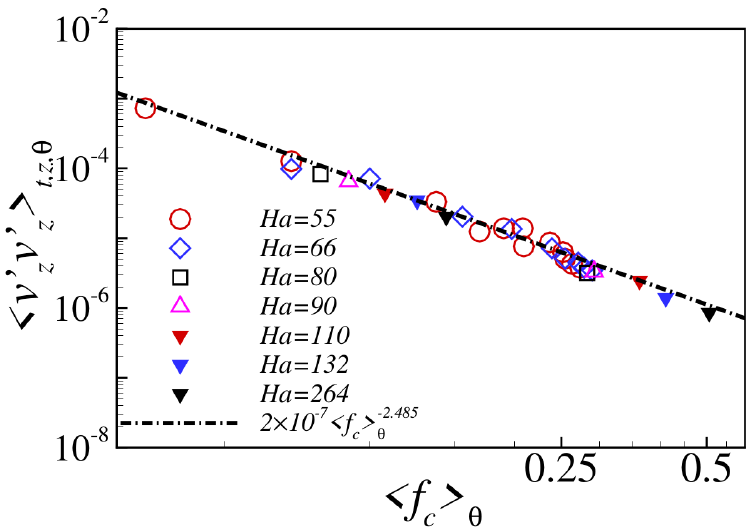}
  \caption{\label{fcUz} The distribution of the average relative turbulent intensity $\langle u^{'}_{z}u^{'}_{z}\rangle_{t,z,\theta}$ with $\langle f_{c}\rangle_{\theta}$, when $Re=15972$, for the different radial values of $\langle f_{c}\rangle_{\theta}$ shown on figure \ref{fcNt}(b).}
\end{figure}
Since the same regions characterise the appearance of three-dimensionality and
three-componentality, the question of how the two are linked naturally arises. To gain insight into it,
we have sought a relationship between the local energy in the fluctuations of the third velocity component $\langle u_z^{\prime2}\rangle_{t,z,\theta}$ and the cutoff frequency, which provides a measure of three-dimensionality across the turbulent spectrum.
These quantities for several cases are plotted on figure \ref{fcUz}.
The collapse of the data into a single curve shows that the degree of the two-dimensionality of the
energy spectrum is tightly linked to the amount of energy in the third component. $\langle u_z^{\prime2}\rangle_{t,z,\theta}$ is therefore solely determined by the true interaction parameter $N_t$ and follows a simple power law, $i.e.$ $\langle u_z^{\prime2}\rangle_{t,z,\theta} =2\times 10^{-7}\langle f_c \rangle_\theta^{-2.485}$. The maximum relative error between the fitting curve and the numerical data is lower than $2.2\%$. As such, as increasing the magnetic field drives the flow towards a quasi-two-dimensional, two-component state, both the transitions to the quasi-two-dimensional state and to the two-component state are progressive and controlled by the true interaction parameters through scalings  $f_c\simeq 0.063 N_t^{0.37}$, and
$\langle u_z^{\prime2}\rangle_{t,z,\theta} \simeq 0.126\times 10^{-7}N_t^{-0.92}$.
It is noteworthy that this scaling implies a different dependence on $Ha$ $\langle u_z^{\prime2}\rangle_{t,z,\theta}\sim Ha^{-1.84}$ than
that for $E^\prime_z= \pi R^2 a\langle \langle u_z^{\prime 2}\rangle_{t,z,\theta}\rangle_r\sim Ha^{-5.76}$. Since the two quantities only differ though radial averaging, the difference can be understood by noticing that if fluctuations are confined to a thin radial region of thickness $\delta_{3D}$ then $E^\prime_z\sim \pi R^2a ({\delta_{3D}}/{R})\langle u_z^{\prime2}\rangle_{t,z,\theta,\delta_{3D}}$, where the subscript $\delta_{3D}$ indicates averaging over that region. Under this assumption, it would follow that {$\delta_{3D}/R\sim Ha^{-3.92}$} (finding the scaling with $Re$ would require extra simulations over a wider range of its values). The corresponding region is much thinner than the free shear layer and the  wall side layer (see section 3.6). However the radial profiles of vertical velocity fluctuations (Fig. \ref{UrzandEtor}) suggest that a sharp peak of vertical velocity fluctuations develops within the later, that would explain this scaling. This suggests that the dominant contribution to the three-component turbulence in the regimes explored in this paper arises out of a very thin region of thickness within the wall side layer.
\section{Conclusions}\label{sec:4}
The present study reports three-dimensional direct numerical simulations of electrically driven MHD turbulent shear flows in
the MATUR experiment \citep{Messadek2002}. The numerical results, which are obtained in a configuration where current is injected far from side walls (at a radius of $r_{e}=5.4$)  and in regimes where the Hartmann layers remain laminar ($18.2\leq R\leq 290.4$), provide
accurate solutions in excellent agreement, not only with the experimental data, but also with the 2D PSM model and other theoretical approaches. Crucially, the DNS, provide access to the
detail of the three-dimensional dynamics. This enabled us to identify the location and the nature of
three-dimensional structures, and the role they play in the overall flow dynamics.

The simulations reproduce typical flow features observed in the experiment and predicted by the theories.
The velocity field is dominated by a limited number of large coherent structures formed at the unstable shear layer. Their dynamics and the inner structure of the layers led us to identify two changes in behaviour when the ratio of two-dimensional inertia to the Lorentz force $R=Re/Ha$ was varied:  from $ R\simeq 121$, small scale turbulence appears in the wall side layer, a value that is consistent with previous findings in other configurations with curved walls \citep{zhao2012}. For $ R\geq 145.2$, that layer separates from the wall, most likely under the influence fluctuations induced by the large vortices in the outer region.
The thickness $\delta_{\rm SW}$ of the wall side layers follows a sequence following these changes of regime: in the laminar regime, it
converges to the asymptotic scaling $\delta_{\rm SW}\sim (Re Ha^{3/2})^{-1}$ theoretically predicted by \cite{Potherat2000}, but becomes much thicker with a scaling of $\delta_{\rm SW}\sim (Re Ha^{-3/2})^{-0.42}$ following the onset of three-dimensional turbulence.

In addition,the energy spectra exhibit a significant dependence on $R$: for $18.2\leq R\leq 60.5$, the turbulent spectrum possesses an inertial range with a $E(f)\sim f^{-3}$ power law. For $ R\geq121$, inertia plays a greater role,  the spectra show a transition frequency between low frequency ranges where $E(f)\sim f^{-5/3}$  and high frequency ranges $E(f)\sim f^{-3}$, similar to the split between inverse energy and direct enstrophy cascades found in quasi-2D MHD flows by \cite{Sommeria1986}.
The dynamics of the free shear layer seen in the experiments was also recovered as DNS showed that the thickness of the free shear layer varies nearly as the vortex size does (scaling as $R^{1/2.3}$ and $R_{L}^{1/2.4}$, respectively).

Detailed analysis of the secondary flow confirmed the phenomenology identified by \cite{Potherat2000,Potherat2005} whereby for $N\lesssim1$, the global recirculation associated to the main azimuthal flow induces a flux of angular momentum towards the wall side layer. The ensuing
thinning of that layer is responsible for an increased dissipation. The intensity of the recirculation matches closely the PSM prediction. At statistical equilibrium, the energy associated to average axial flow component was found to scale as $E_{z}^m\sim Ha^{-5.58}$ both in PSM and the DNS, confirming that it is driven by this main recirculation. The energy associated to the fluctuating part of
this component was on the other hand underestimated by PSM compared to the DNS for $N\lesssim1$.
The discrepancy originates in the small scale turbulence produced in the wall side layer. This effect was found to incur a significant increase in wall shear stress and in turn a reduction in the global angular momentum, a phenomenon that previous theories could not capture. The extra dissipation is also a possible cause for the shortened transient time observed when the
flow is initiated at rest.

A major benefit of the 3D DNS was to afford a detailed scrutiny of the
three-dimensional effects.
The main source of three-dimensionality was found in the small-scale turbulence fed by the mean shear in the free shear layer and the wall side layer.
Outside of these regions, turbulent spectra still exhibit a high frequency range of three-dimensional
structures and a low frequency range of quasi-two dimensional ones, as predicted by \cite{Sommeria1982}
and found in turbulence driven by a crystal of vortices \citep{KLEIN2010}. As in this case, the frequency separating these two ranges scales with the true interaction parameter $N_t$, albeit with a different exponent as $f\sim N^{0.37}$ instead of $f\sim N^{2/3}$. The difference is due to the presence of a strong background flow and when converted into wavenumbers, both experiments exhibit the same cutoff scaling of
$k_c\sim N_t^{1/3}$ \citep{Baker2018}, as predicted by \cite{Sommeria1982}. This results shows that the existence of a cutoff wavelength separating three and quasi-two dimensional turbulent fluctuations extends to sheared turbulence. As such, this result and the associated scaling can be expected to hold in a wider class of flows including duct flows.

Simultaneous access to all three velocity components further enabled us to establish a link between the local dimensionality of the turbulence (measured by the cutoff frequency $f_c$) and its componentality, measured by the energy in the axial component of the velocity fluctuations. While the latter decreases monotonically with the former, and both are controlled by the true interaction parameter
through simple power laws $\langle u_z^{\prime2}\rangle_{t,z,\theta} \sim 0.126 N_t^{-0.92}$ and $\langle f_c \rangle_{\theta}\simeq 0.063 N_t^{0.37}$. Unlike for the dimensionality scaling ($f_c$), no prediction exists for the dimensionality scaling ($\langle u_z^{\prime2}\rangle_{t,z,\theta}$), so this raises the question of its applicability to other types of quasi-static MHD turbulence, beyond shear flow turbulence or even beyond this particular experiment. The question is all the more relevant as in the regime explored here, three-component turbulence was found almost exclusively within a very thin region of the wall side layer. As such, a further step in understanding the link between componentality and dimensionality could target flows where three-dimensional turbulence is more broadly distributed.\\


The authors acknowledge the support from NSFC under Grants $\sharp$51636009, $\sharp$51606183 and CAS under Grants $\sharp$XDB22040201, $\sharp$QYZDJ-SSW-SLH014.
The authors also greatly appreciate the anonymous reviewers for their comments, which significantly helped to improve the manuscript.\\

Declaration of Interests. The authors report no conflict of interest.


\begin{thebibliography}{31}
\expandafter\ifx\csname natexlab\endcsname\relax\def\natexlab#1{#1}\fi
\def\au#1{#1} \def\ed#1{#1} \def\yr#1{#1}\def\at#1{#1}\def\jt#1{\textit{#1}}
  \def\bt#1{#1}\def\bvol#1{\textbf{#1}} \def\vol#1{#1} \def\pg#1{#1}
  \def\publ#1{#1}\def\arxiv#1{#1}\def\org#1{#1}\def\st#1{\textit{#1}}

\bibitem[Alboussi{\`e}re {\em et~al.\/}(1999)Alboussi{\`e}re, Uspenski \&
  Moreau]{Alboussiere1999}
{\sc \au{Alboussi{\`e}re, T}, \au{Uspenski, V} \& \au{Moreau, R}} \yr{1999}
  \at{Quasi-two-dimensional mhd turbulent shear layers}.  \jt{Experimental
  Thermal and Fluid Science}  \bvol{20},  \pg{19--24}.

\bibitem[Alexakis \& Biferale(2018)]{alexakis2018_pr}
{\sc \au{Alexakis, A.} \& \au{Biferale, L.}} \yr{2018}  \at{Cascades and
  transitions in turbulent flows}.  \jt{Physics Reports}  \bvol{767-769},
  \pg{1–101}.

\bibitem[Baker {\em et~al.\/}(2015)Baker, Poth{\'e}rat \&
  Davoust]{Potherat2015}
{\sc \au{Baker, Nathaniel~T}, \au{Poth{\'e}rat, Alban} \& \au{Davoust,
  Laurent}} \yr{2015}  \at{Dimensionality, secondary flows and helicity in
  low-rm mhd vortices}.  \jt{Journal of Fluid Mechanics}  \bvol{779},
  \pg{325--350}.

\bibitem[Baker {\em et~al.\/}(2018)Baker, Poth{\'e}rat, Davoust \&
  Debray]{Baker2018}
{\sc \au{Baker, Nathaniel~T}, \au{Poth{\'e}rat, Alban}, \au{Davoust, Laurent}
  \& \au{Debray, Fran{\c{c}}ois}} \yr{2018}  \at{Inverse and direct energy
  cascades in three-dimensional magnetohydrodynamic turbulence at low magnetic
  reynolds number}.  \jt{Physical review letters}  \bvol{120}~(22),
  \pg{224502}.

\bibitem[Davidson(1997)]{Davidson1997}
{\sc \au{Davidson, PA}} \yr{1997}  \at{The role of angular momentum in the
  magnetic damping of turbulence}.  \jt{Journal of Fluid Mechanics}
  \bvol{336},  \pg{123--150}.

\bibitem[Davidson \& Poth{\'e}rat(2002)]{Davidson2002}
{\sc \au{Davidson, PA} \& \au{Poth{\'e}rat, A}} \yr{2002}  \at{A note on
  b{\"o}dewadt--hartmann layers}.  \jt{European Journal of Mechanics-B/Fluids}
  \bvol{21}~(5),  \pg{545--559}.

\bibitem[Eckert {\em et~al.\/}(2001)Eckert, Gerbeth, Witke \&
  Langenbrunner]{Eckert2001}
{\sc \au{Eckert, S}, \au{Gerbeth, G}, \au{Witke, W} \& \au{Langenbrunner, H}}
  \yr{2001}  \at{Mhd turbulence measurements in a sodium channel flow exposed
  to a transverse magnetic field}.  \jt{International journal of heat and fluid
  flow}  \bvol{22}~(3),  \pg{358--364}.

\bibitem[Klein \& Poth{\'e}rat(2010)]{KLEIN2010}
{\sc \au{Klein, R} \& \au{Poth{\'e}rat, Alban}} \yr{2010}  \at{Appearance of
  three dimensionality in wall-bounded mhd flows}.  \jt{Physical review
  letters}  \bvol{104}~(3),  \pg{034502}.

\bibitem[Kljukin \& Kolesnikov(1989)]{Kljukin1989}
{\sc \au{Kljukin, A, A} \& \au{Kolesnikov, Yu.~B}} \yr{1989} {\em Liquid Metal
  Magnetohydrodynamics\/}, ,  \vol{vol.~10}.  \publ{Kluwer}.

\bibitem[Kobayashi(2006)]{Kobayashia2006}
{\sc \au{Kobayashi, Hiromichi}} \yr{2006}  \at{Large eddy simulation of
  magnetohydrodynamic turbulent channel flows with local subgrid-scale model
  based on coherent structures}.  \jt{Physics of Fluids}  \bvol{18}~(4),
  \pg{045107}.

\bibitem[Kobayashi(2008)]{Kobayashia2008}
{\sc \au{Kobayashi, Hiromichi}} \yr{2008}  \at{Large eddy simulation of
  magnetohydrodynamic turbulent duct flows}.  \jt{Physics of Fluids}
  \bvol{20}~(1),  \pg{015102}.

\bibitem[Kolesnikov \& Tsinober(1974)]{Kolesnikov1974}
{\sc \au{Kolesnikov, Yu~B} \& \au{Tsinober, AB}} \yr{1974}  \at{Experimental
  investigation of two-dimensional turbulence behind a grid}.  \jt{Fluid
  Dynamics}  \bvol{9}~(4),  \pg{621--624}.

\bibitem[Messadek \& Moreau(2002)]{Messadek2002}
{\sc \au{Messadek, Karim} \& \au{Moreau, Rene}} \yr{2002}  \at{An experimental
  investigation of MHD quasi-two-dimensional turbulent shear flows}.
  \jt{Journal of Fluid Mechanics}  \bvol{456},  \pg{137--159}.

\bibitem[Moffatt(1967)]{moffatt1967_jfm}
{\sc \au{Moffatt, H.~K.}} \yr{1967}  \at{On the suppression of turbulence by a
  uniform magnetic field}.  \jt{Journal of Fluid Mechanics}  \bvol{28}~(3),
  \pg{571–592}.

\bibitem[Moresco \& Alboussi{\`e}re(2003)]{Moresco2003}
{\sc \au{Moresco, P} \& \au{Alboussi{\`e}re, T}} \yr{2003}  \at{Weakly
  nonlinear stability of Hartmann boundary layers}.  \jt{European Journal of
  Mechanics-B/Fluids}  \bvol{22}~(4),  \pg{345--353}.

\bibitem[Moresco \& Alboussiere(2004)]{Moresco2004}
{\sc \au{Moresco, Pablo} \& \au{Alboussiere, Thierry}} \yr{2004}
  \at{Experimental study of the instability of the hartmann layer}.
  \jt{Journal of Fluid Mechanics}  \bvol{504},  \pg{167--181}.

\bibitem[Ni {\em et~al.\/}(2007)Ni, Munipalli, Huang, Morley \& Abdou]{Ni2007}
{\sc \au{Ni, Ming-Jiu}, \au{Munipalli, Ramakanth}, \au{Huang, Peter},
  \au{Morley, Neil~B} \& \au{Abdou, Mohamed~A}} \yr{2007}  \at{A current
  density conservative scheme for incompressible MHD flows at a low magnetic
  Reynolds number. part ii: On an arbitrary collocated mesh}.  \jt{Journal of
  Computational Physics}  \bvol{227}~(1),  \pg{205--228}.

\bibitem[Poth{\'e}rat \& Dymkou(2010)]{Potherat2010}
{\sc \au{Poth{\'e}rat, Alban} \& \au{Dymkou, Vitali}} \yr{2010}  \at{Direct
  numerical simulations of low-$Rm$ MHD turbulence based on the least dissipative
  modes}.  \jt{Journal of Fluid Mechanics}  \bvol{655},  \pg{174--197}.

\bibitem[Poth{\'e}rat \& Klein(2014)]{Potherat2014}
{\sc \au{Poth{\'e}rat, Alban} \& \au{Klein, Rico}} \yr{2014}  \at{Why, how and
  when MHD turbulence at low Rm becomes three-dimensional}.  \jt{Journal of
  Fluid Mechanics}  \bvol{761},  \pg{168--205}.

\bibitem[Poth\'erat \& Kornet(2015)]{pk2015_jfm}
{\sc \au{Poth\'erat, A.} \& \au{Kornet, K.}} \yr{2015}  \at{The decay of
  wall-bounded MHD turbulence between walls, at low $Rm$}.  \jt{Journal of
  Fluid Mechanics}  \bvol{683},  \pg{605--636}.

\bibitem[Poth{\'e}rat \& Schweitzer(2011)]{Potherat2011}
{\sc \au{Poth{\'e}rat, Alban} \& \au{Schweitzer, Jean-Philippe}} \yr{2011}
  \at{A shallow water model for magnetohydrodynamic flows with turbulent
  Hartmann layers}.  \jt{Physics of Fluids}  \bvol{23}~(5),  \pg{055108}.

\bibitem[Poth{\'e}rat {\em et~al.\/}(2000)Poth{\'e}rat, Sommeria \&
  Moreau]{Potherat2000}
{\sc \au{Poth{\'e}rat, A}, \au{Sommeria, J} \& \au{Moreau, R}} \yr{2000}
  \at{An effective two-dimensional model for mhd flows with transverse magnetic
  field}.  \jt{Journal of Fluid Mechanics}  \bvol{424},  \pg{75--100}.

\bibitem[Poth{\'e}rat {\em et~al.\/}(2005)Poth{\'e}rat, Sommeria \&
  Moreau]{Potherat2005}
{\sc \au{Poth{\'e}rat, Alban}, \au{Sommeria, Jo{\"e}l} \& \au{Moreau,
  Ren{\'e}}} \yr{2005}  \at{Numerical simulations of an effective
  two-dimensional model for flows with a transverse magnetic field}.
  \jt{Journal of Fluid Mechanics}  \bvol{534},  \pg{115--143}.

\bibitem[Roberts(1967)]{Roberts1967}
{\sc \au{Roberts, Paul~Harry}} \yr{1967} {\em An introduction to
  magnetohydrodynamics\/}, ,  \vol{vol.~6}.  \publ{Longmans London}.

\bibitem[Schumann(1976)]{schumann1976_jfm}
{\sc \au{Schumann, U.}} \yr{1976}  \at{Numerical simulation of the transition
  from three- to two-dimensional turbulence under a uniform magnetic field}.
  \jt{Journal of Fluid Mechanics}  \bvol{74}~(1),  \pg{31–58}.

\bibitem[Sommeria(1986)]{Sommeria1986}
{\sc \au{Sommeria, Joel}} \yr{1986}  \at{Experimental study of the
  two-dimensional inverse energy cascade in a square box}.  \jt{Journal of
  Fluid Mechanics}  \bvol{170},  \pg{139--168}.

\bibitem[Sommeria \& Moreau(1982)]{Sommeria1982}
{\sc \au{Sommeria, Jo{\"e}L} \& \au{Moreau, Ren{\'e}}} \yr{1982}  \at{Why, how,
  and when, MHD turbulence becomes two-dimensional}.  \jt{Journal of Fluid
  Mechanics}  \bvol{118},  \pg{507--518}.

\bibitem[Stelzer {\em et~al.\/}(2015{\natexlab{{\em a\/}}})Stelzer, C{\'e}bron,
  Miralles, Vantieghem, Noir, Scarfe \& Jackson]{Stelzer2015a}
{\sc \au{Stelzer, Zacharias}, \au{C{\'e}bron, David}, \au{Miralles, Sophie},
  \au{Vantieghem, Stijn}, \au{Noir, J{\'e}r{\^o}me}, \au{Scarfe, Peter} \&
  \au{Jackson, Andrew}} \yr{2015{\natexlab{{\em a\/}}}}  \at{Experimental and
  numerical study of electrically driven magnetohydrodynamic flow in a modified
  cylindrical annulus. i. base flow}.  \jt{Physics of Fluids}  \bvol{27}~(7),
  \pg{077101}.

\bibitem[Stelzer {\em et~al.\/}(2015{\natexlab{{\em b\/}}})Stelzer, Miralles,
  C{\'e}bron, Noir, Vantieghem \& Jackson]{Stelzer2015b}
{\sc \au{Stelzer, Zacharias}, \au{Miralles, Sophie}, \au{C{\'e}bron, David},
  \au{Noir, J{\'e}r{\^o}me}, \au{Vantieghem, Stijn} \& \au{Jackson, Andrew}}
  \yr{2015{\natexlab{{\em b\/}}}}  \at{Experimental and numerical study of
  electrically driven magnetohydrodynamic flow in a modified cylindrical
  annulus. ii. instabilities}.  \jt{Physics of Fluids}  \bvol{27}~(8),
  \pg{084108}.

\bibitem[Tabeling \& Chabrerie(1981)]{Tabeling1981}
{\sc \au{Tabeling, P} \& \au{Chabrerie, JP}} \yr{1981}  \at{Magnetohydrodynamic
  Taylor vortex flow under a transverse pressure gradient}.  \jt{Physics of
  Fluids}  \bvol{24}~(3),  \pg{406--412}.

\bibitem[Zhao \& Zikanov(2012)]{zhao2012}
{\sc \au{Zhao, Yurong} \& \au{Zikanov, Oleg}} \yr{2012}  \at{Instabilities and
  turbulence in magnetohydrodynamic flow in a toroidal duct prior to transition
  in Hartmann layers}.  \jt{Journal of Fluid Mechanics}  \bvol{692},
  \pg{288--316}.

\end{thebibliography}

                                                                                                                          183,1         Bot

\end{document}